\documentclass{article}

\usepackage{arxiv}

\usepackage[utf8]{inputenc} 
\usepackage[T1]{fontenc}    
\usepackage{natbib}
\usepackage[hidelinks]{hyperref}       
\usepackage{booktabs}       
\usepackage{amsfonts}       
\usepackage{amsmath}
\usepackage{nicefrac}       
\usepackage{microtype}      
\usepackage{graphicx}
\usepackage{array}
\usepackage{tabularx}
\usepackage{tikz}
\usetikzlibrary{arrows.meta,positioning,calc,backgrounds,fit}
\graphicspath{{./figures/}}

\definecolor{pnavy}{HTML}{1F4E79}
\definecolor{pllm}{HTML}{C0392B}   
\definecolor{pemb}{HTML}{2F7D5B}   
\definecolor{prul}{HTML}{2B6CB0}   
\tikzset{
  pipebase/.style={rounded corners=2.5pt, align=center, inner sep=5.5pt, font=\footnotesize, line width=0.7pt},
  dat/.style={pipebase, draw=black!45, fill=black!7},
  llm/.style={pipebase, draw=pllm!80, fill=pllm!10},
  emb/.style={pipebase, draw=pemb!85, fill=pemb!12},
  rul/.style={pipebase, draw=prul!85, fill=prul!10},
  outp/.style={pipebase, draw=pnavy, fill=pnavy!14, line width=1pt},
  annot/.style={font=\scriptsize, text=black!60, align=left},
  flow/.style={-{Stealth[length=2.6mm]}, line width=0.9pt, draw=black!55},
  reuse/.style={-{Stealth[length=2.4mm]}, line width=0.8pt, draw=black!45, dashed},
  stepnum/.style={circle, fill=pnavy, text=white, font=\bfseries\scriptsize, inner sep=1.6pt, minimum size=4mm},
  lanehead/.style={font=\scriptsize\bfseries, text=black!55, anchor=south west},
  lanebox/.style={rounded corners=4pt, draw=black!18, fill=black!2}
}
\newcommand{\pipelegend}{%
  \begin{tikzpicture}[baseline=-2pt]
    \node[dat, minimum height=3.2mm, minimum width=6mm, inner sep=1.5pt] (l1) {};
    \node[annot, right=1.2mm of l1] (t1) {data / artifact};
    \node[llm, minimum height=3.2mm, minimum width=6mm, inner sep=1.5pt, right=4mm of t1] (l2) {};
    \node[annot, right=1.2mm of l2] (t2) {LLM judgment (cached)};
    \node[emb, minimum height=3.2mm, minimum width=6mm, inner sep=1.5pt, right=4mm of t2] (l3) {};
    \node[annot, right=1.2mm of l3] (t3) {embedding};
    \node[rul, minimum height=3.2mm, minimum width=6mm, inner sep=1.5pt, right=4mm of t3] (l4) {};
    \node[annot, right=1.2mm of l4] (t4) {deterministic rules};
    \node[outp, minimum height=3.2mm, minimum width=6mm, inner sep=1.5pt, right=4mm of t4] (l5) {};
    \node[annot, right=1.2mm of l5] {output};
  \end{tikzpicture}}

\title{The Pulse Beneath the Job Title:\\ Monthly Readings of Requirements and Tasks\\ from 750 Million Chinese Job Ads}

\author{
 Qin Chen \\
  Shanghai DeepPulse Technology\\
  \texttt{qinchen1986@hotmail.com} \\
 \And
 Ying Fang \\
  Wang Yanan Institute for Studies in Economics (WISE)\\
  and School of Economics\\
  and Laboratory of Digital Finance\\
  Xiamen University\\
  \texttt{yifst1@xmu.edu.cn} \\
   \And
 Xiangyu Wang \\
 Wang Yanan Institute for Studies in Economics (WISE)\\
  Xiamen University\\
  \texttt{wangxiangyu@stu.xmu.edu.cn} \\
  \And
 Leo Yang Yang \\
 Department of Accountancy, Economics and Finance \\
  School of Business \\
  Hong Kong Baptist University\\
  \texttt{leoyang@hkbu.edu.hk} \\
}

\begin{document}
\maketitle

\begin{abstract}
How do we define an occupation? By its job title? An accountant at a small trading company keeps the books; at a listed firm the same title demands a certified-accountant licence, and the week goes to the reports that regulators and the board read. Same title, different bar, different work. What defines an occupation is who it lets in and what it asks them to do. In a rapidly changing labor market, tracking those requirements and tasks is how to take the market's pulse. Yet no instrument reads both at the speed they change. Official occupational directories like O*NET report one national average per occupation, updated every few years. Job postings are timely but unstructured. Research built on them works from job titles plus proprietary skill keywords, which blur what is asked \emph{of} a candidate into what a candidate is asked \emph{to do}. The blur matters, because rising requirements and changing tasks are different events with different causes. We separate them. From 752.6~million job ads posted on China's five leading recruitment platforms between 2022 and 2026, we \emph{extract} the phrases employers write, \emph{unify} those that name the same thing, and \emph{validate} the mapping from text back to entry. By doing so we construct two catalogs, 20{,}721 requirements a candidate must meet and 44{,}479 tasks the hire will do. With the entries standardized, we annotate them further. Each task, for example, carries a score for how far a language model could absorb it. Matched back onto every ad, the catalogs read the market month by month. Two examples show what the layer beneath the job title buys. First, the occupational registry records one accountant where the ads record a staircase, the junior certificate at the bottom of the wage range and the intermediate one at the top. Second, counting occupations says the work most exposed to language models is disappearing, and counting tasks says far less of it is. The tasks outlive the jobs that carried them, and the door into the better-paid ones demands the harder certificate.
\end{abstract}


\vspace{2pt}\noindent\textbf{JEL classification:} J23, J24, O33, C81, C55.

\clearpage

\section{Introduction}
\label{sec:intro}

Ask a small trading company what its accountant does, and the answer is recording receipts, reconciling accounts, closing the month. Ask a listed firm, and the answer is tax planning, the reports regulators require, and the analysis that goes to the board; hardly a single duty overlaps. Then ask each what it takes to get the job. The first wants someone ``careful and willing to learn''. The second wants the certified-accountant license and says so before it mentions anything else. Through all of this, one thing stays fixed. The business card and the statistical registry both say accountant. Same title, different bar, different work. What, then, does define an occupation?

Not the few words of the title, but who it lets in and what it asks them to do. Neither half is our invention. For two decades, the task approach in labor economics has understood occupations through what is done in them, with technology acting on tasks, substituting for some and complementing others, and reaching occupations and wages only through how tasks are recombined \citep{autor2003skill,acemoglu2011tasks,autor2013task}. The other half, the requirements that govern access to a posted job, has its own literature, in which what an employer demands is not a property of the job but a screen the employer sets and resets: requirements rise when workers are plentiful and fall when they are scarce, within the same firm and the same job title \citep{modestino2020}. Tasks define the job; requirements decide who reaches it. Nor are titles empty: title words alone predict a posting's pay and its flow of applicants \citep{marinescu2020}. The problem is that the granularity stops there. A record keyed to the title is blind in two ways. First, it collapses each occupation to a point, when an occupation is really a distribution over postings: the same title spans very different jobs and very different bars, and when the market moves, it is largely the weights of this distribution that shift, so both the spread and its movement vanish the moment postings are averaged into one national profile. Second, it cannot follow tasks across jobs: a task's presence in the market is defined by every posting that mentions it, not by any single occupation's headcount, so when an occupation contracts, its tasks live on elsewhere, on a margin invisible to a record that keeps only the title or attaches one fixed task profile to it. Theory long ago pushed the definition of an occupation below the title; what has been missing is measurement to match.

Neither of the two instruments economists use to read labor demand supplies that measurement. Official occupational directories have structure but not frequency. For a quarter century the reference has been O*NET, the U.S.\ Department of Labor's decomposition of occupations into tasks and work activities on one side and worker and experience requirements, licensing included, on the other \citep{peterson2001onet,nrc2010onet}; the empirical task literature runs on it, and nearly every influential measure of technology exposure is built on top of it, from early automation probabilities to the latest LLM-exposure ratings \citep{frey2017future,brynjolfsson2018,webb2020,felten2021,eloundou2024,handa2025,hampole2025}. But its production process, questionnaires aggregated to one national profile per occupation, fixes both cadence and resolution. A full update cycle takes years: the task list for computer programmers changed little from the 2015 release to the 2026 release, while the actual work of programmers was rewritten more than once. And within an occupation there is, by construction, no variation to study, so exposure research has had to treat the occupation as an atom. China's official occupational classification shares the architecture, without task statements at all \citep{mohrss2022}. Job postings have the opposite half: frequency without structure. Vacancy text is by now a standard demand-side lens, for skill upgrading in recessions \citep{hershbein2018}, skill differences across firms \citep{deming2018}, and AI hiring \citep{acemoglu2022vacancies}; but its working unit remains the job title plus proprietary skill keywords \citep{carnevale2014}, which blur what is asked \emph{of} a candidate into what a candidate is asked \emph{to do}. Commercial taxonomies are black boxes; the classic task treatment of posting text classifies it into O*NET's own vocabulary \citep{atalay2020} rather than deriving a vocabulary from employer language; Europe's monitoring system codes ads into a curated expert thesaurus \citep{cedefop2022,esco2022}. And for Chinese-language postings, whose screening content is unusually rich and institutionally consequential (gender preferences: \citealp{kuhn2013gender}; age limits: \citealp{helleseter2020}; the effects of banning them: \citealp{kuhn2023}), no open vocabulary of either kind exists at all. Structure without frequency, or frequency without structure; the language that most needs the vocabulary is the one without it.

That both sides move is not a conjecture; it is among the best-documented facts in this literature. Using a century of newspaper job ads, \citet{atalay2020} show that the task content of American work changed substantially, and that most of the change happened within occupations, inside titles that never changed. \citet{autor2024newfrontiers} show that more than six in ten of today's jobs are in titles that did not exist in 1940: technology keeps minting new tasks and retiring old ones, rewriting the occupational directory itself. The current shock fits the pattern: from occupation-level automation forecasts \citep{frey2017future} to AI hiring effects visible in postings \citep{acemoglu2022vacancies}, the evidence says change is underway. The question is at which layer it shows up first. Employment is a quasi-fixed factor: dismissing and replacing workers is costly \citep{oi1962labor,hamermesh1993labor}, so an employer whose needs shift need not begin by adjusting its existing workforce; it may instead alter the occupations it recruits or the tasks bundled within them, leaving the data to show which of these margins moves first. We take this ordering as motivation, not as a claim we test. History says the content moves; adjustment costs make recomposing work one of the cheaper margins to move first. Our own window says the same on the other side: from late 2023 onward, the requirements Chinese employers posted rose as credential gates spread. In a rapidly changing labor market, then, tracking those requirements and tasks is how to take the market's pulse, and the pulse has to be taken monthly, not on the next survey round.

A missing instrument would matter less in a small or slow-moving market. China's is neither. Its roughly 734~million workers are one in five on earth \citep{nbs2025communique}: whatever happens to work here happens to more people than anywhere else. And the market moves: an economy that compressed a century of structural change into two generations, where occupations appear, swell, and shrink at a pace mature economies rarely see. The timing sharpens the stakes further: the development and deployment of AI is now a story of two countries, and China is one of them. On the American side of that story, the instruments for watching labor demand exist: O*NET, payroll records, posting archives. On the Chinese side, they largely do not. A shock of historic scale is meeting the world's largest labor market, and no instrument is watching.

The challenge, then, is concrete: turn the requirements and tasks in job-ad text into stable, countable units, apply those units across hundreds of millions of advertisements, and keep the resulting record inexpensive to update month after month. None of the three demands is trivial. Employers write ``maintain client relationships'' in hundreds of ways and ``communication skills'' in dozens, and every one of them must land in the same entry, while ``equipment maintenance'' and ``client maintenance,'' one word apart, must never merge. The data run to hundreds of millions of ads and over a hundred million distinct descriptions, so anything priced per document, whether human or model, is ruled out. And the market coins new phrases every month, so the vocabulary must be cheap to reuse and cheap to re-scan, or the instrument is obsolete the day it is finished. Only when all three hold does taking the market's pulse mean anything.

This paper takes up that challenge. The data are 752.6~million job ads posted on China's five leading recruitment platforms between January 2022 and June 2026, de-duplicated to 116~million unique job descriptions. We distill employers' own language into a standardized vocabulary in three moves. A large language model first \emph{extracts} atomic phrases from a 2.31-million-description sample stratified by occupation, one action, ability, credential, or condition per phrase. The phrases are then \emph{unified}: embeddings cluster them, a model adjudicates roughly 520{,}000 boundary pairs (does this phrasing name the same thing as that one?), and connected components merge; ``maintain client relationships'' alone absorbs 797 surface variants, and annotation of this kind matches or beats crowd workers \citep{gilardi2023}. The mapping from free text to standard entry is finally \emph{validated}: candidate regular expressions are generated for every entry and scored, entry by entry, against the sampled benchmark; the best of five is kept, failures are dropped, and the vocabulary is frozen. The frozen vocabulary is then matched back onto every ad. It has two halves, and keeping them apart is part of the design: 20{,}721 requirements record what the candidate must be or hold, and 44{,}479 tasks record what the hire will do, a distinction that proprietary skill keywords conflate and that matters most in a market where licenses, degrees, identity gates, and physical requirements do much of the screening. One cost discipline runs through the construction: expensive judgment is confined to a bounded extraction sample and to reusable artifacts, such as category crosswalks, cluster boundaries, taxonomy assignments, and matching patterns, rather than being applied posting by posting. Adjudicating all vocabulary boundaries costs about \yen3{,}300; naively classifying titles with the full occupational directory in context would cost roughly \yen9.7 million per 100 million titles.

The result is a dynamic O*NET for China: 12.1~billion item--posting mentions across the full 2022--2026 record, observable month by month and sliceable by occupation, city, wage, education, and employer type. O*NET does what an occupational information system is meant to do: provide a vocabulary comparable across occupations. Dynamic, because the production function has changed: the vocabulary comes from employer text rather than questionnaires, its readings update monthly rather than every few years, and they arrive as distributions rather than one national average; and once the vocabulary is frozen, the full five-year history can be re-scanned in hours, so every number in this paper is measured with one method rather than a patchwork of vintages. And a dynamic system can do things a static one cannot. Score each entry once on every dimension defined for its side: eleven of thirteen for tasks and nine for requirements, covering how the work relates to machines, people, places, and institutions; the scores ride onto every ad that mentions it: what a static system publishes as one national number per occupation, frozen between survey rounds, becomes a monthly reading, sliceable by city, wage, or employer, and re-scorable as the technology itself evolves. Exposure to large language models, the dimension this paper leans on most, is measured exactly this way.

Two examples show what the layer beneath the job title buys. First, one occupation code turns out to cover a staircase. Core bookkeeping accounts for 68\% of an \emph{accountant's} task mentions in the lowest wage band and 49\% in the highest, displaced by compliance and data analysis. The bar moves with it, and in two registers at once: the share of postings demanding some certificate nearly doubles across the lower bands and then saturates, while inside the class, the entry-level title gives way, with the intermediate one rising from 1.9\% of postings to a peak of 14.1\%. \emph{Production worker} is more extreme on the task side: the bundle's LLM exposure is about 0.17 in the lowest wage band and about 0.48 in the highest, where the ads describe quality documentation and process coordination, not assembly. The counterexample is just as informative: \emph{Java developer} carries nearly the same bundle at every experience level, on both sides at once, and its average can be trusted. Certificates barely enter at all: whatever the experience level, only 3--5\% of Java postings ask for one. Which occupations a single average summarizes well is itself a question only within-occupation data can answer. Pulled wide, the record charts this variation title by title, across market segments, month by month; and since late 2023, the task composition of posted work has drifted away from what language models absorb and toward hands-on, on-site, higher-stakes work, a drift that a fixed-weight counterfactual attributes chiefly to re-weighting across occupations rather than rewriting within them.

Second, the record lets us look beneath a pattern now being documented. Payroll studies in the United States have begun to find employment falling fastest in the most AI-exposed occupations \citep{brynjolfsson2025canaries}. Going one layer down starts with a definition: in our record, an occupation is exposed exactly to the degree that its tasks are; the occupation is nothing but a weighted average of its tasks. That distinction opens two parallel descriptive gradients in the data. Occupations built mostly of exposed tasks are losing posting share sharply (slope $-1.33$); the broad task categories they are built of lose mention share far more slowly ($-0.83$). A shift-share decomposition locates the gap: the task-level gradient runs almost entirely through the posting-share channel; on average, exposed tasks lose ground because the occupations that carry them shrink, while within surviving occupations the task mix stays flat, with no systematic stripping of exposed content. Here, the layer question above meets its answer: within posted demand, adjustment runs through which occupations are posted, not through what the surviving ones contain. The tasks outlive the jobs that carried them. Both gradients are descriptive, as consistent with cyclical stories as with technological ones; but seeing them at all requires measuring the same data at two layers at once.

This study contributes to three strands of literature. For occupational information systems, it changes how occupational information is produced: employer text, monthly updates, and distributions. For posting-based measurement, it supplies what proprietary taxonomies have not: an open, documented vocabulary for Chinese-language postings, one that separates what a candidate must \emph{be} (requirements) from what the hire will \emph{do} (tasks). To the measurement of technology and work, it measures technology exposure from the task content of job postings at monthly frequency; the exposure scores converge with the leading published measures at $r=0.73$--$0.79$, so exposure and realization can be read from one instrument, complementing the realization evidence emerging from payrolls \citep{brynjolfsson2025canaries}, freelancing platforms \citep{hui2024}, and productivity experiments \citep{noy2023,brynjolfsson2025genai}. Section~\ref{sec:data} describes the data. Section~\ref{sec:pipeline} documents the construction stage by stage. Section~\ref{sec:validation} collects the validation evidence. Section~\ref{sec:facts} develops the two examples. Section~\ref{sec:access} states how versions are fixed and what the instrument cannot do; Section~\ref{sec:conclusion} concludes.

\section{Data}
\label{sec:data}

Chinese employers advertise vacancies on large commercial recruitment platforms, where an employer opens an account, posts an advertisement for each opening, and screens applicants through the site. What the platform records is the advertisement. We observe the five leading platforms, anonymized as platforms A--E under our data-use agreements,\footnote{The marginals we publish (Tables~\ref{tab:marginals} and~\ref{tab:representativeness}) may allow industry insiders to guess at platform identities; the anonymization follows the agreements' terms rather than claiming concealment.} from January 2022 through June 2026, roughly 1.8~terabytes of text. Each advertisement carries a free-text title, a description laying out tasks and requirements, and a set of structured fields: city, posted wage range, required experience and education, employer size, industry tags, and, where stated, college-major requirements. The unit of observation throughout is the job ad.

\begin{figure}[t]
\centering
\includegraphics[width=0.72\textwidth]{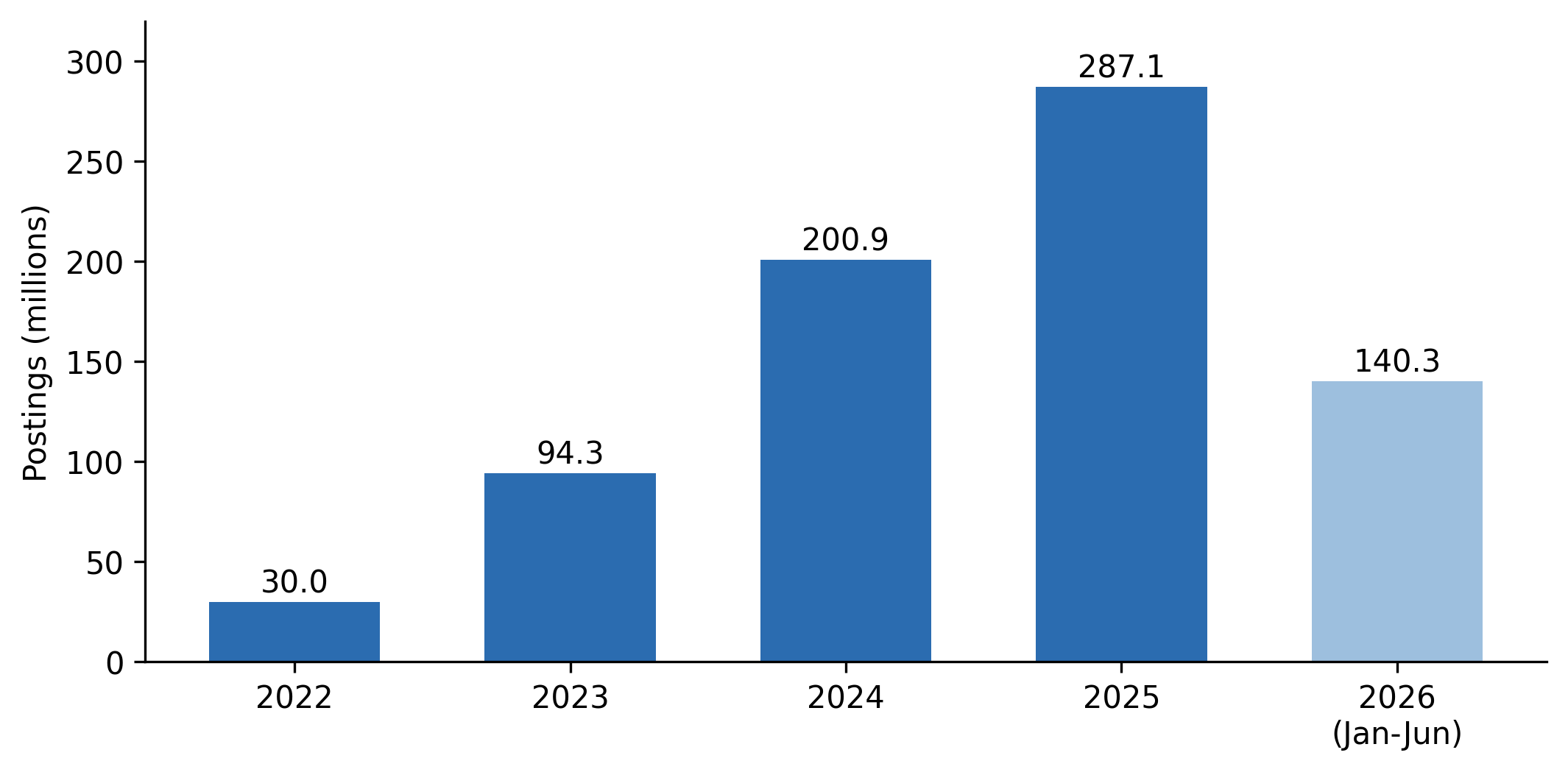}
\caption{Postings by year, five platforms, 752.6~million in total. The 2026 bar covers January--June; at the first-half run rate, full-year 2026 is on pace with 2025.}
\label{fig:corpus}
\end{figure}

Figure~\ref{fig:corpus} shows the data by year: 30~million postings in 2022, 287~million in 2025, and 752.6~million in total, with all five platforms active in every year of the window. Part of that growth is real, and part reflects expanding platform coverage; the two cannot be fully separated. Every analysis in this paper that compares over time therefore uses within-platform shares or restricts itself to platform subsets with stable coverage, never raw counts. Within-platform shares hold the platform mix fixed but do not guard against selection moving within a platform, which is why the per-platform coverage series reported with the validation evidence accompany every temporal claim.

The same description text recurs across cities, months, and platforms, so we separate two objects. Text processing runs once per unique description: after markup, numbering, whitespace, and punctuation are normalized away, exact-duplicate bodies collapse to 116~million unique descriptions, and every judgment about a text, extraction, unification, the regular-expression scan, is made once and applies to all its copies. Statistics run at the posting level: each description's matches are joined back to every posting that carries it, together with that posting's own structured fields, so a description posted in thirty cities counts thirty times, each with its own city, wage, and employer. Weighting by posting is an economic choice, not a bookkeeping one: each posting is a separate offer, addressed to one city's searchers in one month at a stated wage, so the posting-weighted composition is the demand that job seekers actually face, and the within-title results of Section~\ref{sec:facts} live on precisely this variation, the same text meeting different wages in different markets. A re-post need not be a new vacancy, so posting weights read as advertised demand rather than as vacancy counts; weighting by unique description instead is the natural robustness convention and matters mainly for high-churn segments such as staffing-agency hiring. The results below are not artifacts of the convention: counting each unique description once steepens the occupation-layer gradient of Section~\ref{sec:twolayers} from $-1.33$ to $-1.76$ (one vote per employer, $-1.43$), and the aggregate exposure drift of Figure~\ref{fig:dim13} is $-0.013$ under one-posting-one-vote against $-0.012$ mention-weighted.

The structured fields do the slicing in everything downstream, so their coverage matters. Coverage is high but not universal, and analyses that condition on a field inherit its selection. A posted wage range is present on 97.4\% of postings. An explicit education requirement appears on 65.8\% and an explicit experience band on 50.8\%; the remainders are dominated not by missing fields but by ads that affirmatively state \emph{no} requirement (``education unrestricted,'' ``experience unrestricted''), which we retain as their own categories rather than treat as missing. Table~\ref{tab:marginals} summarizes the data, the marginal distribution of postings across wage band, education, experience, employer size, and platform, on which every slice in Section~\ref{sec:facts} conditions. One accounting rule matters throughout: structured fields attach to postings, not to de-duplicated description texts, since one unique description recurs across cities, wage offers, and platforms.

\begin{table}[t]
\centering
\caption{Summary of the data: marginal distribution of the 752.6~million postings, percent.}
\label{tab:marginals}
\footnotesize
\begin{tabular}{lr@{\hskip 2.4em}lr}
\toprule
\multicolumn{2}{l}{\textbf{Posted monthly wage}} & \multicolumn{2}{l}{\textbf{Employer size (employees)}} \\
\midrule
Below \yen5{,}000 & 10.0 & Fewer than 50 & 16.2 \\
\yen5{,}000--8{,}000 & 32.6 & 50--100 & 18.1 \\
\yen8{,}000--12{,}000 & 27.6 & 100--500 & 19.3 \\
\yen12{,}000--20{,}000 & 17.1 & 500--1{,}000 & 9.5 \\
\yen20{,}000--30{,}000 & 6.5 & More than 1{,}000 & 21.6 \\
\yen30{,}000--50{,}000 & 2.9 & Not stated & 15.4 \\
\yen50{,}000 and above & 0.7 & & \\
No posted wage & 2.6 & & \\
\midrule
\multicolumn{2}{l}{\textbf{Education requirement}} & \multicolumn{2}{l}{\textbf{Experience requirement}} \\
\midrule
Unrestricted (none stated) & 34.2 & Unrestricted & 49.2 \\
Associate degree or below & 37.5 & Under 1 year & 20.8 \\
Bachelor's degree & 26.5 & 1--3 years & 19.0 \\
Graduate degree & 1.7 & 3--5 years & 9.0 \\
 & & 5 years or more & 2.0 \\
\midrule
\multicolumn{4}{l}{\textbf{Platform:}\quad A 34.6 \quad B 20.0 \quad C 19.7 \quad D 15.7 \quad E 10.0} \\
\bottomrule
\end{tabular}
\end{table}

The data have two limitations. First, online postings over-represent urban, formal, white-collar and service hiring relative to census employment. Table~\ref{tab:representativeness} quantifies the tilt against the 2020 population census: sales and service work is 49.6\% of postings against 33.9\% of census employment, professionals 21.0\% against 10.4\%, while agriculture (20.5\% of employment) is essentially absent from the platforms; by industry, information \& software (13.0\% vs.\ 1.7\%) and leasing \& business services (13.2\% vs.\ 2.7\%) are heavily over-represented while construction (2.0\% vs.\ 11.3\%) and wholesale \& retail are under-represented. The table's bottom panel adds the geographic margin of the same tilt: tier-1 and tier-2 cities carry 54.8\% of postings against 20.0\% of census employment, while tier-6 localities (49.2\% of census employment) supply 19.7\% of postings. Second, a posting is a vacancy advertisement, not a hire. The system measures the composition of expressed demand, the object of interest for studying how employers rewrite work, and any level statement about the aggregate labor market must be read through the selection in Table~\ref{tab:representativeness}.

\begin{table}[t]
\centering
\caption{Representativeness of the data: posting shares (2022--2026, five platforms) vs.\ 2020 census employment shares, percent. City tiers follow the standard six-tier classification of Chinese cities; census tier shares aggregate employment by prefecture. The industry panel shows selected sections; rows need not sum to 100.}
\label{tab:representativeness}
\footnotesize
\begin{tabular}{lrr@{\hskip 1.8em}lrr}
\toprule
\textbf{Occupation class} & \textbf{Census} & \textbf{Postings} & \textbf{Industry section} & \textbf{Census} & \textbf{Postings} \\
\midrule
Sales \& service workers & 33.9 & 49.6 & Agriculture & 20.6 & 0.0 \\
Production \& manufacturing & 25.8 & 19.9 & Manufacturing & 18.1 & 26.4 \\
Agriculture, forestry \& fishing & 20.5 & 0.0 & Wholesale \& retail & 14.1 & 6.0 \\
Professionals \& technicians & 10.4 & 21.0 & Construction & 11.3 & 2.0 \\
Clerical staff & 6.9 & 8.2 & Transport \& logistics & 5.0 & 1.4 \\
Managers & 2.2 & 1.4 & Hotels \& catering & 4.9 & 1.2 \\
Other / unclassified & 0.2 & 0.0 & Education & 4.1 & 2.4 \\
 & & & Leasing \& business services & 2.7 & 13.2 \\
 & & & Information \& software & 1.7 & 13.0 \\
 & & & Finance & 1.5 & 3.2 \\
\midrule
\textbf{City tier} & \textbf{Census} & \textbf{Postings} & \textbf{City tier} & \textbf{Census} & \textbf{Postings} \\
\midrule
Tier 1 & 6.5 & 21.0 & Tier 5 & 5.8 & 3.1 \\
Tier 2 & 13.5 & 33.8 & Tier 6 & 49.2 & 19.7 \\
Tier 3 & 11.3 & 15.6 & Other / unmatched & 4.5 & 1.0 \\
Tier 4 & 9.2 & 5.8 & & & \\
\bottomrule
\end{tabular}
\end{table}

\section{Measurement}
\label{sec:pipeline}
\label{sec:related}

Survey-based occupational infrastructure answers the question ``what is this occupation, on average, in this country.'' The question technology shocks now pose is ``what is this occupation \emph{becoming}, this month, in this city, at this wage level,'' and answering it takes a different production function. O*NET succeeded the Dictionary of Occupational Titles in 1998 and remains the reference decomposition of work into tasks, detailed work activities, skills, knowledge, and abilities \citep{peterson2001onet}; its strengths, cross-occupation comparability and three decades of continuity, come from the same survey production process that fixes its two limits of cadence and resolution, discussed in the introduction \citep{nrc2010onet}. Europe's ESCO shares the architecture (a curated multilingual taxonomy of occupations, skills, and competences) and the limits \citep{esco2022}. China's official classification enumerates roughly 1{,}600 occupations with prose descriptions, revised at multi-year intervals, without task statements \citep{mohrss2022}. None of the three can be read at the frequency at which the market now moves.

The raw material for the missing answer already exists. Vacancy text is a standard demand-side lens \citep{carnevale2014,hershbein2018,deming2018,acemoglu2022vacancies}, extended across a century by historical text \citep{atalay2020,autor2024newfrontiers}. What is missing is not data but vocabulary: open, documented units in which employer text can be counted. Proprietary skill keywords conflate what a worker must \emph{be} (requirements) with what the job \emph{is} (tasks), objects that behave differently throughout Section~\ref{sec:facts}. The closest institutional precedent, Cedefop's Skills-OVATE, continuously codes European vacancies \emph{into} the curated ESCO vocabulary \citep{cedefop2022,esco2022}, and the direction is the problem: a fixed expert taxonomy cannot register work the market has just invented. For Chinese-language postings no open vocabulary of either kind existed at all, despite a literature showing how much screening content Chinese ads carry: explicit gender preferences \citep{kuhn2013gender}, age limits and their interaction with gender \citep{helleseter2020}, the effects of banning such preferences \citep{kuhn2023}, plus the certification, physical, household-registration, and work-context requirements documented below.

The stakes are clearest in the measurement of AI exposure. Exposure measures have progressed from occupation-level automation probabilities \citep{frey2017future} to task- and ability-level scores aggregated up \citep{brynjolfsson2018,webb2020,felten2021,eloundou2024}, and most recently to usage-based measures \citep{handa2025} and firm-level aggregation \citep{hampole2025}; realized impacts have begun to be read on payrolls \citep{brynjolfsson2025canaries}, freelancing platforms \citep{hui2024}, and in productivity experiments \citep{noy2023,brynjolfsson2025genai}. Every exposure measure inherits the cadence and resolution of the catalog underneath it. Move the catalog to the demand side at market frequency, and exposure becomes a property of a posted task mix, observable monthly, per occupation, per city, per wage band, so exposure and realization can be read from the same instrument, and the within-occupation margin becomes measurable for the first time in a Chinese context.

These gaps dictate the construction. The vocabulary must be \emph{derived from} employer text rather than imposed on it, so that the taxonomy is an output of the data, not a filter on it, and is re-derivable when the market invents new work. It must keep requirements and tasks apart, with explicit room for the screening content above, which then becomes countable: how often employers demand a specific license, an age band, or shift-pattern tolerance, monthly, by market segment. And it must refresh monthly at the scale of hundreds of millions of ads, which rules out spending model judgment on individual postings. The affordable design spends LLM judgment only on \emph{reusable artifacts}, drawing on the annotation literature in which LLMs match or beat crowd workers \citep{gilardi2023} while answering its known fragilities: every judgment, a crosswalk entry, a cluster boundary, a taxonomy cell, a regular expression, is made once under a recorded prompt version, benchmarked before deployment (the grader exam of Section~\ref{sec:scores}, the per-bucket pattern benchmarks of Section~\ref{sec:backfill}), treating the model as a measurement device to be calibrated rather than an oracle, and then replayed deterministically, so model drift cannot silently re-label history and any prompt revision triggers an explicit, versioned re-adjudication. What results borrows O*NET's central idea, a reusable vocabulary comparable across occupations, and changes the production function: the raw material is the employer's own text, the update cycle is monthly, and the unit of description is the posting rather than the occupation, so occupation profiles become distributions that vary over time, space, and market segment rather than single points. Table~\ref{tab:compare} summarizes the contrast; the two production functions are complements, not substitutes, O*NET's curated three-decade comparability being irreplaceable.

\begin{table}[t]
\centering
\caption{Survey-based vs.\ posting-based occupational infrastructure.}
\label{tab:compare}
\small
\begin{tabularx}{\textwidth}{lXX}
\toprule
 & \textbf{O*NET (U.S.)} & \textbf{This system (China)} \\
\midrule
Unit of description & Occupation-average profile & Posting-level task mix, aggregated on demand \\
Source & Incumbent/analyst questionnaires & 752.6M employer-written ads \\
Update cycle & Multi-year survey rounds & Monthly; full re-scan in hours \\
Vocabulary & 19{,}281 tasks; 2{,}087 DWAs; 332 IWAs & 65{,}200 items: 20{,}721 requirements (16 classes, 181 subclasses); 44{,}479 tasks ($32\times26$ grid) \\
Within-occupation variation & None by construction & By month, city, industry, wage, education, experience, firm size \\
Technology scores & Periodic expert supplements & 13 continuous dimensions per item, re-scorable on demand \\
Validation & Survey methodology & Per-bucket pattern benchmarks; external convergence with published exposure measures \\
\bottomrule
\end{tabularx}
\end{table}

Each nonstandard input gets the standardization its form demands. Occupations and industries come first (Section~\ref{sec:coord}): they are the coordinates along which everything downstream is sampled, sliced, and validated. The pair of catalogs is the core (Section~\ref{sec:reqres}): employers' own phrases are extracted, unified into a shared vocabulary, and organized into its two halves, the 20{,}721 requirements, among them the college-major field, the one requirement standardized against an official catalog rather than derived from employer language, and the 44{,}479 tasks. Scoring adds thirteen derived dimensions to the finished items (Section~\ref{sec:scores}), and the backfill matches the finished vocabulary back onto every ad, where validation lives (Section~\ref{sec:backfill}). Throughout, embeddings handle retrieval and grouping, and compiled regular expressions handle the hundreds of millions of documents; LLM judgment enters only through the cached artifacts just described. Appendix~\ref{app:models} records the model behind every LLM-marked node, and the choices have a logic: where a judgment shapes results it was selected by benchmark, the deployed grader by a fourteen-model exam (Section~\ref{sec:scores}) and the regular expressions by keeping the best of five generators per bucket (Section~\ref{sec:backfill}), while elsewhere the model of record is documented so that every judgment can be traced.

\subsection{The occupation and industry codes}
\label{sec:coord}

Both fields arrive as platform vocabularies, category trees for occupations and industry lists, and both are standardized the same way: a crosswalk judged once per entry, cached, and inherited by every posting that carries the entry.

\subsubsection{Occupations}
\label{sec:occ}

Everything downstream begins by placing each posting in a standard occupation, and the raw material is the title, the noisiest field on the ad: Chinese titles interleave occupation with industry, location, perks, and slogans (``Love to win---sales rep---five insurances,'' ``Health-checkup client manager (weekends off)''). No off-the-shelf design survives this at scale. Embedding nearest neighbors track industry vocabulary rather than occupational core, filing the health-checkup client manager next to clinical-laboratory physicians instead of sales agents. Models asked to emit official codes from memory fabricate or mismatch them for 80--100\% of titles across six frontier models, a failure of the task rather than of any model. And the two designs accurate enough to trust, supplying the full occupational catalog in context or retrieval-augmented selection, price per title, roughly \yen9.7~million per 100~million titles, at prohibitive latency.

\begin{figure}[t]
\centering
\resizebox{\textwidth}{!}{%
\begin{tikzpicture}
\def\cA{0}\def\cB{5.9}\def\cC{11.8}
\begin{scope}[on background layer]
\draw[lanebox] (-2.45,1.55) rectangle (14.25,-3.45);
\draw[lanebox] (-2.45,-3.95) rectangle (14.25,-8.10);
\end{scope}
\node[lanehead, anchor=north west] at (-2.30,1.47) {PHASE I · BUILD ONCE: expensive judgments cached as reusable artifacts};
\node[lanehead, anchor=north west] at (-2.30,-4.03) {PHASE II · APPLY TO EVERY POSTING: near-zero marginal cost};
\node[dat, text width=4.3cm] (a1) at (\cA,0)
  {Platform job taxonomies\\ \scriptsize 5 platforms; thousands of leaf categories\\ \scriptsize e.g.\ ``Internet/AI $>$ back-end $>$ .NET''};
\node[llm, text width=4.3cm] (a2) at (\cB,0)
  {LLM category crosswalk\\ \scriptsize each tree node judged once};
\node[dat, text width=4.3cm] (a3) at (\cC,0)
  {Category crosswalk table\\ \scriptsize O*NET-SOC $\cdot$ \mbox{CN-2022}\\ \scriptsize unified (1{,}267)};
\node[dat, text width=4.3cm] (b1) at (\cA,-2.50)
  {Reference-period ads\\ \scriptsize titles \emph{with} platform categories};
\node[rul, text width=4.3cm] (b2) at (\cC,-2.50)
  {Title dictionary\\ \scriptsize title $\to$ platform category $\to$ standard occupation};
\draw[flow] (a1) -- (a2);
\draw[flow] (a2) -- (a3);
\draw[flow] (a3) -- (b2);
\draw[flow] (b1) -- (b2) node[annot, midway, above=1.2mm, align=center] {every (title, category)\\ pair becomes a row};
\node[dat, text width=4.3cm] (c1) at (\cA,-5.00)
  {New title, \emph{has} platform category};
\node[rul, text width=4.3cm] (c2) at (\cB,-5.00)
  {Exact dictionary lookup};
\node[dat, text width=4.3cm] (c3) at (\cA,-7.10)
  {New title, \emph{unseen}, no category};
\node[emb, text width=4.3cm] (c4) at (\cB,-7.10)
  {Embedding retrieval:\\ \scriptsize nearest dictionary-labeled titles vote};
\node[outp, text width=4.3cm] (d) at (\cC,-6.05)
  {Every posting mapped into 3 occupation systems\\ \scriptsize O*NET-SOC $\cdot$ \mbox{CN-2022}\\ \scriptsize unified (1{,}267)};
\draw[flow] (c1) -- (c2);
\draw[flow] (c3) -- (c4);
\draw[flow] (c2.east) -- ($(d.west)+(0,0.55)$);
\draw[flow] (c4.east) -- ($(d.west)+(0,-0.55)$);
\draw[reuse] (b2.south) -| (c2.north) node[annot, pos=0.78, right=1.2mm] {artifact reuse};
\node[stepnum] at ($(a1.north west)+(0.05,0)$) {1};
\node[stepnum] at ($(a2.north west)+(0.05,0)$) {2};
\node[stepnum] at ($(b2.north west)+(0.05,0)$) {3};
\node[stepnum] at ($(c2.north west)+(0.05,0)$) {4a};
\node[stepnum] at ($(c4.north west)+(0.05,0)$) {4b};
\node[stepnum] at ($(d.north west)+(0.05,0)$) {5};
\end{tikzpicture}}

\vspace{2pt}
\pipelegend
\caption{Occupation standardization, in two phases. Phase~I (top lane) spends the expensive judgments once: an LLM crosswalks each platform's category tree to the three target systems, and reference-period ads turn that crosswalk into a title dictionary. Phase~II (bottom lane) classifies the incoming stream at near-zero marginal cost, by exact lookup where a platform category exists and by embedding retrieval where it does not. Colors encode the engine throughout the paper: red = cached LLM judgment, green = embedding, blue = deterministic rules, gray = data artifacts, navy = output. A worked example: the platform category ``Internet/AI $>$ back-end development $>$ .NET'' is judged once, mapping to O*NET-SOC 15-1252 (software developers), its counterpart in the 2022 national classification, and the back-end-development category of the working taxonomy; every title arriving under that category, ``senior .NET engineer,'' ``C\# back-end,'' inherits the mapping by lookup, and only a title with no platform category at all is classified by the vote of its embedded neighbors. Section~\ref{sec:occ} gives the design comparison behind this choice.}
\label{fig:pipe-occ}
\end{figure}
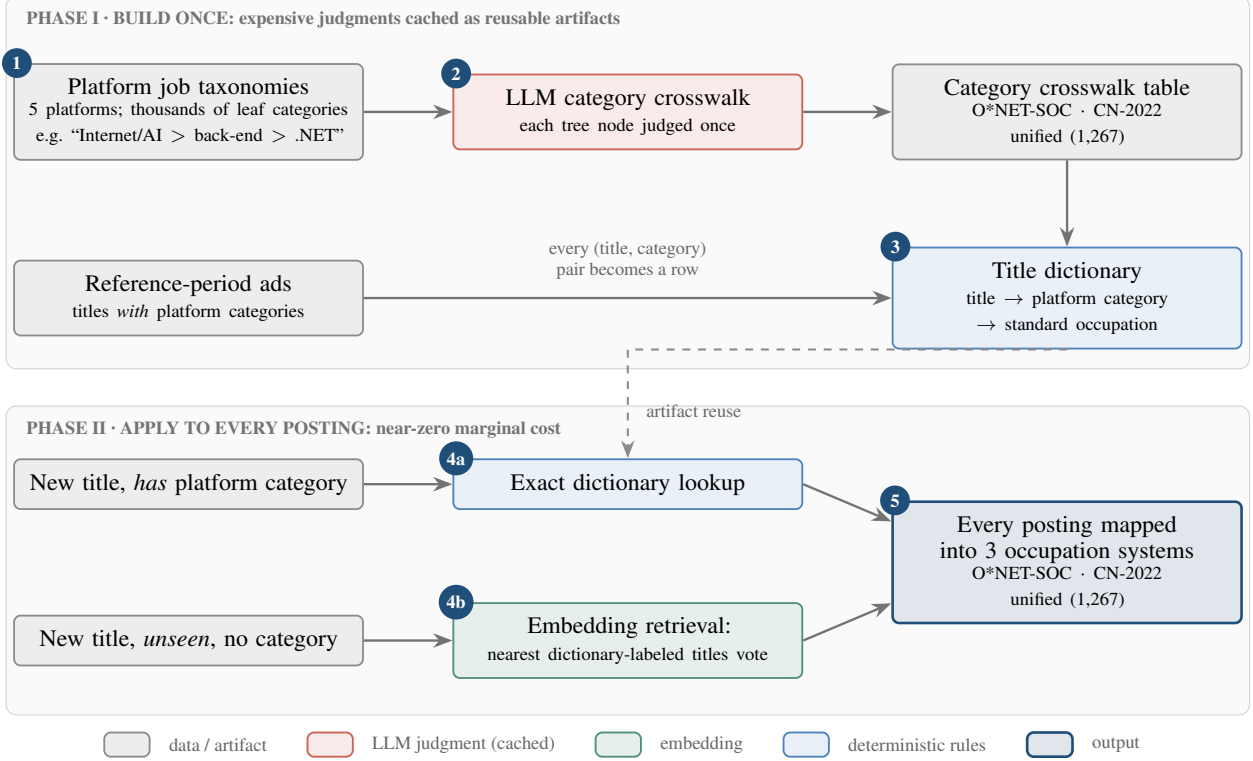

The production method inverts the problem: \emph{classify categories, not titles}. The platforms themselves maintain detailed internal job taxonomies (e.g., ``Internet/AI $>$ back-end development $>$ .NET''), and the overwhelming majority of ads carry a platform category. We crosswalk each platform's category tree (thousands of nodes, not hundreds of millions of titles) to three target systems with a high-capability LLM: O*NET-SOC, China's official 2022 classification \citep{mohrss2022}, and a unified working taxonomy of 1{,}267 occupation categories designed to be stable across platforms. The judgment attaches to the category and is made once per node; every observed (title, platform category) pair then becomes a row in a cached lookup table by deterministic join of that crosswalk, not by a separate judgment. A title never seen before classifies through its category alone. Only when the category itself is absent does the assignment fall back to embedding retrieval against the labeled titles, where the nearest vote. New months therefore classify at near-zero marginal cost, and the lookup table improves monotonically as adjudicated pairs accumulate. Figure~\ref{fig:pipe-occ} lays out the two phases; the color convention introduced there (red for cached LLM judgments, green for embedding operations, blue for deterministic rules) is used in all stage diagrams below.

\subsubsection{Industries}
\label{sec:ind}

\begin{figure}[t]
\centering
\resizebox{\textwidth}{!}{%
\begin{tikzpicture}
\def\cA{0}\def\cB{5.9}\def\cC{11.8}
\node[dat, text width=4.3cm] (i1) at (\cA,0)
  {Platform industry vocabularies\\ \scriptsize tags $+$ free text, per platform};
\node[llm, text width=4.3cm] (i2) at (\cB,0)
  {LLM crosswalk\\ \scriptsize each vocabulary entry $\to$ division, judged once};
\node[dat, text width=4.3cm] (i3) at (\cC,0)
  {Industry crosswalk table};
\node[dat, text width=4.3cm] (i4) at (\cA,-2.20)
  {Employers with tagged postings};
\node[rul, text width=4.3cm] (i5) at (\cB,-2.20)
  {Company-name lookup\\ \scriptsize modal division per employer};
\node[outp, text width=4.3cm] (i6) at (\cC,-2.20)
  {96 two-digit divisions\\ \scriptsize on every posting};
\draw[flow] (i1) -- (i2);
\draw[flow] (i2) -- (i3);
\draw[flow] (i4) -- (i5);
\draw[flow] (i5) -- (i6);
\draw[flow] (i3) -- (i6);
\node[stepnum] at ($(i1.north west)+(0.05,0)$) {1};
\node[stepnum] at ($(i2.north west)+(0.05,0)$) {2};
\node[stepnum] at ($(i5.north west)+(0.05,0)$) {3};
\end{tikzpicture}}
\caption{Industry standardization. The LLM crosswalk from platform vocabularies to the 96 national two-digit divisions is judged once (steps 1--2); a company-name lookup table fills in employers whose postings carry no usable tag (step 3). Colors as in Figure~\ref{fig:pipe-occ}.}
\label{fig:pipe-ind}
\end{figure}
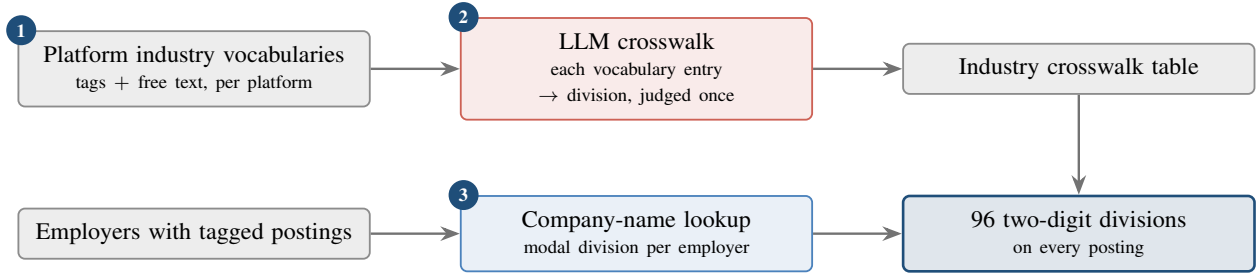

Employer industry tags mix platform-specific vocabularies with free text. We standardize to the two-digit divisions (96 categories) of China's national industry classification in two moves (Figure~\ref{fig:pipe-ind}): an LLM crosswalk from each platform's industry vocabulary to divisions, and a company-name lookup table that assigns divisions to employers whose postings lack usable tags, built from the modal division of each company's tagged postings. Postings that still lack an assignment inherit the division of the same employer name elsewhere in the data.

\subsection{The requirement and task catalogs}
\label{sec:reqres}

\begin{figure}[t]
\centering
\begin{tikzpicture}
\node[llm, text width=6.3cm] (s1) at (0,0)
  {\textbf{Extract}\; LLM atomic-phrase extraction\\ \scriptsize one action / ability / credential / condition per phrase; benefits, boilerplate, structured fields excluded; majors routed to their own field};
\node[rul, text width=6.3cm] (s2) at (0,-1.85)
  {\textbf{Count}\; normalize and count phrases};
\node[emb, text width=6.3cm] (s3) at (0,-3.30)
  {\textbf{Group}\; embed $\to$ PCA $\to$ initial clusters};
\node[llm, text width=6.3cm] (s4) at (0,-5.05)
  {\textbf{Adjudicate}\; embedding recalls candidate neighbors; LLM rules \textsc{same} / \textsc{loose} / \textsc{no} on each pair\\ \scriptsize decided on core nouns and verbs; generic filler ignored};
\node[rul, text width=6.3cm] (s5) at (0,-6.90)
  {\textbf{Merge}\; connected components\\ over \textsc{same}/\textsc{loose} edges};
\node[llm, text width=6.3cm] (s6) at (0,-8.55)
  {\textbf{Organize}\; LLM-assisted taxonomy assignment};
\node[outp, text width=6.3cm] (s7) at (0,-10.30)
  {\textbf{Catalog}\; 65{,}200 items = 20{,}721 requirements + 44{,}479 tasks};
\foreach \a/\b in {s1/s2, s2/s3, s3/s4, s4/s5, s5/s6, s6/s7}{\draw[flow] (\a) -- (\b);}
\node[annot, anchor=west] at (3.6,0)     {sample stratified by occupation\\ over 116M unique job descriptions};
\node[annot, anchor=west] at (3.6,-1.85) {millions of distinct natural phrases,\\ frequency-weighted};
\node[annot, anchor=west] at (3.6,-3.30) {262{,}144 clusters, deliberately\\ over-segmented};
\node[annot, anchor=west] at (3.6,-5.05) {$\sim$520{,}000 pairwise judgments $\approx$ \yen3{,}300;\\ cost scales with the catalog, not the data};
\node[annot, anchor=west] at (3.6,-6.90) {``maintain client relationships'' absorbs\\ 18 clusters and 797 surface variants};
\node[annot, anchor=west] at (3.6,-8.55) {requirements: 16 classes, 181 subclasses;\\ tasks: 32 domains $\times$ 26 actions (698 cells)};
\node[annot, anchor=west] at (3.6,-10.30) {vs.\ O*NET 30.3: 19{,}281 tasks,\\ 2{,}087 DWAs, 332 IWAs};
\foreach \s/\n in {s1/1, s2/2, s3/3, s4/4, s5/5, s6/6}{\node[stepnum] at ($(\s.north west)+(0.05,0)$) {\n};}
\end{tikzpicture}
\caption{The requirement/task catalog in six steps, with the scale reached at each stage in the right margin. Read the color column for where the cost sits: the only LLM expenditure at the scale of the data is step~1's sampled extraction, while steps~4 and~6 spend judgment on category boundaries, whose number grows with the catalog rather than the data. Colors as in Figure~\ref{fig:pipe-occ}; details in Section~\ref{sec:reqres}.}
\label{fig:pipe-reqres}
\end{figure}
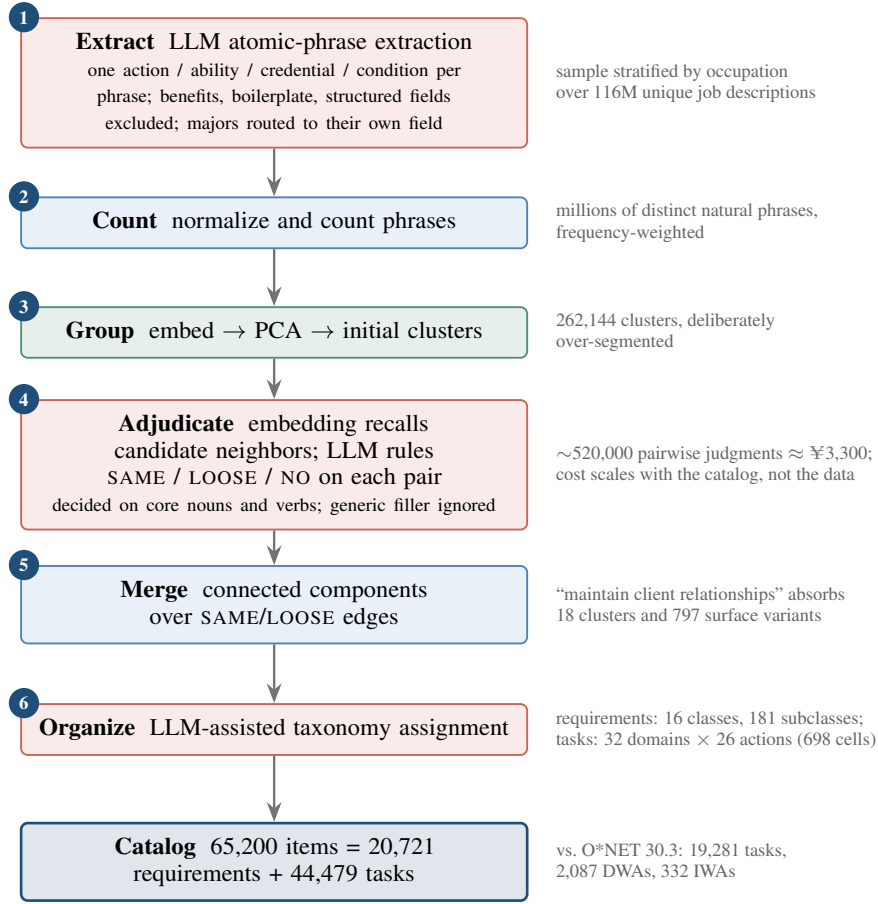

The core of the system converts free-text job descriptions into two countable objects: \emph{requirements} (conditions the candidate must satisfy) and \emph{tasks} (things the hire will do). The distinction matters economically (requirements price the worker; tasks define the job) and technically (they need different taxonomies). Construction has six steps, summarized in Figure~\ref{fig:pipe-reqres}: step~1 performs the extraction of the introduction's vocabulary, steps~2--5 the unification, and step~6 organizes each half into its taxonomy, developed below; validation arrives with the backfill (Section~\ref{sec:backfill}).

\textbf{Step 1: sampled atomic extraction.} From the 116~million unique descriptions we draw a sample of 2.31~million, stratified by occupation and capped at 100 postings per occupation, platform, and year, with within-stratum selection by deterministic hash order so the draw is reproducible; because the caps flatten large occupations, sample frequencies are within-sample counts rather than incidence estimates for the full data, which come from the backfill. Each sampled description is fed to an LLM extractor with a strict contract: return atomic phrases only (one action, ability, credential, or condition per phrase; long sentences are split), preserve meaning while stripping filler, ignore benefits, addresses, and boilerplate, and route major requirements to their own field rather than the requirement list. Structured fields (education, experience, wage, city) are never re-extracted as text. The 2.31-million-description sample is extracted by the production model of Table~\ref{tab:models}; its phrase-level output defines the benchmark frequencies used for pattern selection and backfill validation below.
\textbf{Step 2: counting.} Extracted phrases are normalized and counted, yielding 2.9~million distinct requirement phrases and 7.3~million distinct task phrases with frequencies: raw material, not yet analyzable. The same underlying item (``communication skills'') appears in dozens of surface forms, while boilerplate (``other tasks assigned by leadership'') masquerades as content.

\textbf{Step 3: embedding and initial clustering.} Phrases are embedded, compressed by principal component analysis (PCA), and grouped into 262{,}144 initial clusters, deliberately over-segmented so that step~4 merges rather than splits.

\textbf{Step 4: LLM boundary adjudication.} For each cluster, embedding retrieval proposes candidate neighbors, and an LLM adjudicates each pair with a three-way contract: \textsc{same} (same core object and action; surface variation only), \textsc{loose} (mergeable for analysis; slightly wider boundary), or \textsc{no} (shares only generic words: ``equipment maintenance'' vs.\ ``client maintenance''). The adjudicator is explicitly instructed to ignore generic verbs (``responsible for,'' ``support'') and to decide on core nouns and verbs. This stage performed roughly 520{,}000 pairwise judgments at a total LLM cost of about \yen3{,}300. It is the single most consequential expenditure in the construction, and the clearest illustration of the budget discipline: the judgments are about \emph{category boundaries}, so their number scales with the catalog, not the data.

\textbf{Step 5: connected-component merge.} \textsc{same}/\textsc{loose} edges are resolved into connected components, producing reusable buckets. (A \emph{bucket} is one catalog item; we use the two terms interchangeably.) ``Maintain client relationships'' (62{,}467 mentions) absorbs eighteen source clusters and 797 surface variants; ``cross-department collaboration'' absorbs fifty-one.

The final catalog holds 20{,}721 requirement buckets and 44{,}479 task buckets: 65{,}200 items. The two names follow O*NET's own vocabulary: what a candidate must hold corresponds to its Worker and Experience Requirements, licensing included, and what the hire does corresponds to its task statements and work activities. For scale, O*NET~30.3 contains 19{,}281 task statements, 2{,}087 detailed work activities (DWAs), and 332 intermediate work activities (IWAs). The comparison is imperfect in an instructive way: O*NET's statements are curated survey prose, while our items are compressed employer language (more numerous chiefly because employer phrasings are finer-grained than curated statements, so the counts are not comparable in information content), frequency-weighted, and re-derivable from scratch when the market invents new work.

\begin{table}[t]
\centering
\caption{Top-level structure of the two taxonomies.}
\label{tab:reqtax}
\footnotesize
\begin{tabularx}{\textwidth}{p{0.47\textwidth}X}
\toprule
\textbf{Requirements: 16 classes (single axis)} & \textbf{Tasks: $32\times26$ grid (two axes)} \\
\midrule
Degree \& identity gates; Licenses \& certifications; Language skills; IT \& programming; Tools \& software; Domain-specific skills; Data \& analytics; Generic soft skills; Business, sales \& customer; Knowledge \& theory; Experience; Disposition \& attitude; Physical \& appearance; Work-context fit; Background \& compliance; Resources \& niche fit
&
\emph{Business domains (32):} client acquisition; client relations \& service; sales \& deals; market research \& PR; growth marketing \& ads; platform \& store operations; product management; design \& content; software \& algorithms; hardware \& electronics; testing \& quality; data analysis; research \& experiments; project delivery; engineering \& construction; manufacturing \& process; equipment \& IT operations; procurement; warehousing \& logistics; accounting \& tax; investment; HR; administration; legal \& IP; compliance \& government affairs; teaching \& student services; medical \& health care; on-site commercial service; personal care; safety \& emergency; driving \& transport; translation \& language.
\emph{Action modes (26):} information gathering; data entry \& processing; analysis \& assessment; planning; design \& solution output; development \& coding; production \& assembly; installation \& deployment; testing \& inspection; execution; monitoring \& patrol; maintenance \& operations; reception \& consultation; prospecting; selling \& promotion; coordination \& reporting; delivery closure; troubleshooting; management \& team leading; training \& mentoring; documentation \& filing; compliance submission; settlement \& payment; optimization \& iteration; risk control \& safety; resource scheduling. \\
\bottomrule
\end{tabularx}
\end{table}

\subsubsection{The requirement catalog}
\label{sec:reqcat}
\label{sec:major}

Requirements form a single-axis hierarchy: 16 top-level classes (Table~\ref{tab:reqtax}) refined into 181 subclasses. The classes reserve explicit space for screening content characteristic of Chinese hiring: licenses and certifications; identity gates; physical, health, and appearance requirements; work-context fit (shifts, travel, on-site adaptability); background and compliance checks.

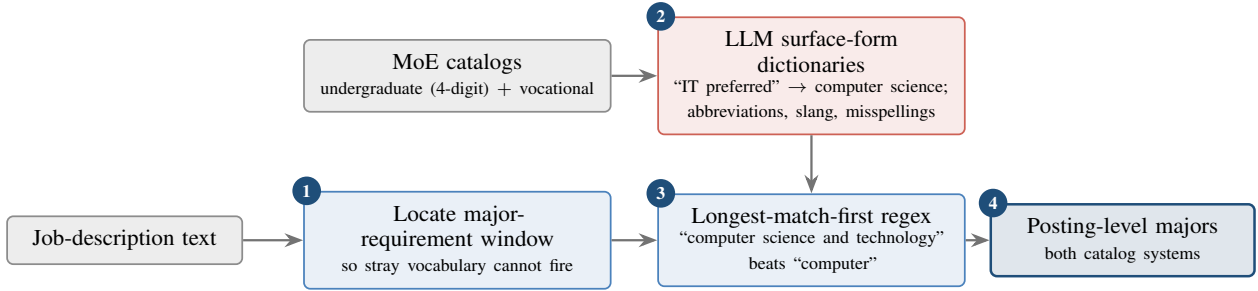
\begin{figure}[t]
\centering
\resizebox{\textwidth}{!}{%
\begin{tikzpicture}
\node[dat, text width=2.9cm] (m1) at (0,0)
  {Job-description text};
\node[rul, text width=3.9cm] (m2) at (4.65,0)
  {Locate major-requirement window\\ \scriptsize so stray vocabulary cannot fire};
\node[dat, text width=3.9cm] (m0) at (4.65,2.30)
  {MoE catalogs\\ \scriptsize undergraduate \mbox{(4-digit)} $+$ vocational};
\node[llm, text width=3.9cm] (m3) at (9.60,2.30)
  {LLM \mbox{surface-form} dictionaries\\ \scriptsize ``IT preferred'' $\to$ computer science;\\ \scriptsize abbreviations, slang, misspellings};
\node[rul, text width=3.9cm] (m4) at (9.60,0)
  {Longest-match-first regex\\ \scriptsize ``computer science and technology''\\ \scriptsize beats ``computer''};
\node[outp, text width=3.3cm] (m5) at (13.95,0)
  {Posting-level majors\\ \scriptsize both catalog systems};
\draw[flow] (m1) -- (m2);
\draw[flow] (m0) -- (m3);
\draw[flow] (m3) -- (m4);
\draw[flow] (m2) -- (m4);
\draw[flow] (m4) -- (m5);
\node[stepnum] at ($(m2.north west)+(0.05,0)$) {1};
\node[stepnum] at ($(m3.north west)+(0.05,0)$) {2};
\node[stepnum] at ($(m4.north west)+(0.05,0)$) {3};
\node[stepnum] at ($(m5.north west)+(0.05,0)$) {4};
\end{tikzpicture}}
\caption{College-major extraction. Window detection confines matching to the span where majors are named (step 1); an LLM turns the official catalogs into dictionaries of employer surface forms (step 2), which compile into longest-match-first regular expressions (step 3) and are backfilled to every posting (step 4). Colors as in Figure~\ref{fig:pipe-occ}.}
\label{fig:pipe-major}
\end{figure}

One requirement type is standardized against an official catalog rather than derived from employer language: the college major. 
Chinese ads frequently name acceptable college majors, but rarely in catalog language: ``IT preferred,'' ``HVAC or industrial-civil construction,'' ``studied nursing.'' Matching proceeds in four steps (Figure~\ref{fig:pipe-major}). (i) A requirement window is located, the sentence span in which majors are named, so that major vocabulary elsewhere in the ad cannot fire. (ii) An LLM reads the Ministry of Education catalogs (the undergraduate catalog at the four-digit class level and the separate vocational catalog) and generates, for every catalog entry, a dictionary of surface forms actually used by employers: full names, abbreviations, colloquialisms, and common misspellings. (iii) The dictionaries are compiled into longest-match-first regular expressions, so ``computer science and technology'' wins over ``computer.'' (iv) Matches are backfilled to every posting, yielding posting-level major requirements in both the undergraduate and vocational systems. The dictionaries are artifacts: reviewing and extending them improves all past and future months simultaneously.

\subsubsection{The task catalog}
\label{sec:taskcat}

Tasks form a two-axis grid: 32 business domains (what object the work acts on: client acquisition, accounting and tax, software and algorithms, manufacturing and process, \ldots) crossed with 26 action modes (how the work is done: research and information gathering, analysis and assessment, development and coding, testing and inspection, coordination and reporting, troubleshooting, \ldots). Of the 832 possible cells, 698 are occupied. Every task bucket therefore answers two questions at once, \emph{what} it operates on and \emph{how}, which is what later makes task re-bundling measurable: bundles are pairs of grid cells.

\subsection{The derived scores}
\label{sec:scores}

Standardization alone does not answer the questions this paper poses. Knowing that a posting mentions \emph{conducting data analysis} says nothing about whether current language models can accelerate or absorb that work, whether it must happen on site, or whether a credential gates entry to it. What standardization changes is the cost of answering. The catalog is finite, 65{,}200 items behind 12.1~billion mentions, so a question that is unanswerable ad by ad becomes a bounded coding exercise: score each item once, and the answer rides onto every mention in the market, monthly, sliceable along the occupation and industry codes. That is the derived layer. Buckets say what the work is; scores say where the work stands relative to machines, people, places, and institutions. Each catalog item carries every continuous 0--1 dimension defined for its side, eleven of the thirteen for tasks and nine for requirements. Seven are defined for every item: \emph{LLM exposure} (can current LLMs/agents substantially accelerate or absorb the core output: text, code, analysis, classification, workflow orchestration); \emph{human complementarity} (dependence on human judgment, trust, relationships, accountability, emotional labor); \emph{cognitive complexity}; \emph{on-site presence}; \emph{field mobility}; \emph{public interaction}; and \emph{ambiguity of phrasing}. LLM exposure deliberately pools acceleration and absorption, two channels with different labor-market implications; the score therefore reads as what models can take on, not as predicted displacement, which is read from the realized outcomes of Section~\ref{sec:twolayers}. Four are meaningful only for tasks and are scored on task items alone: \emph{routine codifiability}, \emph{creative originality}, \emph{error consequences}, and \emph{embodied manual skill}. Two are meaningful only for requirements and are scored on requirement items alone: \emph{digital-skill intensity} and \emph{institutional barriers} (license/credential gates). Aggregations respect this structure: task-only dimensions average over task mentions, requirement-only dimensions over requirement mentions, shared dimensions over both. Side-specific dimensions are stored as null on the inapplicable side and enter no aggregate: every score sum is paired with the event count of its own side, so means are taken only over items on which the dimension is defined.

The grader was selected by examination. Fourteen candidate models scored a common set of 100 buckets against a consensus benchmark; models were then placed on a price--accuracy frontier. The most accurate model was also the most expensive per exam; the cheapest models carried 50--100\% higher error. The selected grader sits on the frontier (second-best accuracy at four-fifths the exam cost of the most accurate, and faster), and scored the full catalog in batch mode. Because scores attach to buckets rather than postings, the entire scoring bill is independent of the size of the data, and re-scoring the catalog under an improved rubric is a bounded, repeatable operation.

Two properties of the scores matter for applications. First, every dimension is a genuine distribution over items, not a binary tag; the anchors are intuitive (frequency-weighted examples: \emph{greeting walk-in customers} scores 0.20 on LLM exposure while \emph{conducting data analysis} scores 0.75; \emph{equipment maintenance} scores 0.95 on on-site presence while \emph{writing technical documentation} scores 0.10; the requirement \emph{CET-4 (College English Test, Band~4) certificate} scores 0.75 on institutional barriers while \emph{communication skills} scores 0.00). Second, the dimensions are not redundant re-labelings of one another. Table~\ref{tab:scores} characterizes all thirteen distributions across the catalog, frequency-weighted by backfilled mentions: every dimension spreads widely (the 10th--90th percentile range is at least 0.30 on each), and the correlation of each dimension with LLM exposure runs from $-0.75$ (embodied manual skill) through $-0.11$ (human complementarity) to $+0.64$ (cognitive complexity). The near-zero correlation with human complementarity is the operative fact for substitution claims: high LLM exposure does not imply low human complementarity (handling customer complaints scores high on both), which is why such claims require reading at least two dimensions jointly.

\begin{table}[t]
\centering
\caption{Score characterization: distributions of the thirteen dimensions across the catalog, frequency-weighted by backfilled mentions. $r$(LLM) is the frequency-weighted correlation with LLM exposure over the items on which the dimension is defined.}
\label{tab:scores}
\footnotesize
\begin{tabular}{llrrrrrr}
\toprule
\textbf{Dimension} & \textbf{Items} & \textbf{Mean} & \textbf{SD} & \textbf{P10} & \textbf{P50} & \textbf{P90} & \textbf{$r$(LLM)} \\
\midrule
LLM exposure & both & 0.44 & 0.23 & 0.10 & 0.45 & 0.75 & --- \\
Human complementarity & both & 0.67 & 0.14 & 0.50 & 0.70 & 0.80 & $-0.11$ \\
Cognitive complexity & both & 0.53 & 0.21 & 0.20 & 0.55 & 0.78 & $+0.64$ \\
On-site presence & both & 0.30 & 0.31 & 0.00 & 0.20 & 0.85 & $-0.28$ \\
Field mobility & both & 0.11 & 0.18 & 0.00 & 0.05 & 0.35 & $-0.14$ \\
Public interaction & both & 0.30 & 0.31 & 0.00 & 0.18 & 0.85 & $+0.04$ \\
Ambiguity of phrasing & both & 0.35 & 0.25 & 0.05 & 0.28 & 0.75 & $-0.31$ \\
Routine codifiability & tasks & 0.54 & 0.18 & 0.30 & 0.52 & 0.80 & $-0.35$ \\
Creative originality & tasks & 0.30 & 0.22 & 0.05 & 0.25 & 0.62 & $+0.46$ \\
Error consequences & tasks & 0.56 & 0.16 & 0.35 & 0.55 & 0.78 & $-0.06$ \\
Embodied manual skill & tasks & 0.18 & 0.24 & 0.00 & 0.05 & 0.60 & $-0.75$ \\
Digital-skill intensity & requirements & 0.14 & 0.26 & 0.00 & 0.00 & 0.60 & $+0.59$ \\
Institutional barriers & requirements & 0.15 & 0.23 & 0.00 & 0.05 & 0.45 & $-0.10$ \\
\bottomrule
\end{tabular}
\end{table}

\subsection{The backfill}
\label{sec:backfill}

\begin{figure}[t]
\centering
\resizebox{\textwidth}{!}{%
\begin{tikzpicture}
\def\cB{5.0}\def\cC{10.0}\def\cD{14.3}
\node[dat, text width=2.9cm] (k0) at (0,-1.25)
  {requirement/task catalog\\ \scriptsize 65{,}200 items};
\node[llm, text width=3.9cm] (k3) at (\cB,0)
  {Regex generation\\ \scriptsize 5 LLMs $\times$ 2--8 patterns per bucket};
\node[rul, text width=3.9cm] (k4) at (\cC,0)
  {Best-of-5 selection\\ \scriptsize per-bucket sampled benchmark;\\ \scriptsize 353k patterns};
\node[rul, text width=3.3cm] (k5) at (\cD,0)
  {Hyperscan scan\\ \scriptsize 116M unique JDs; hours};
\node[llm, text width=3.9cm] (k1) at (\cB,-2.50)
  {Grader exam\\ \scriptsize 14 models $\times$ 100 buckets vs.\\ \scriptsize consensus; price--accuracy frontier};
\node[dat, text width=3.9cm] (k2) at (\cC,-2.50)
  {65{,}200 items scored\\ \scriptsize 11 dims per task, 9 per requirement; batch; bill independent\\ \scriptsize of data size};
\node[outp, text width=6.2cm] (k6) at (12.15,-4.95)
  {12.1B item mentions on 752.6M postings\\ \scriptsize 16.1 per posting $\cdot$ 92 rollup tables $\cdot$ benchmark log-$r$ 0.92};
\draw[flow] (k0.east) -- (k3.west);
\draw[flow] (k0.east) -- (k1.west);
\draw[flow] (k3) -- (k4);
\draw[flow] (k4) -- (k5);
\draw[flow] (k1) -- (k2);
\draw[flow] (k5.south) -- (k5.south |- k6.north);
\draw[flow] (k2.south) -- (k2.south |- k6.north) node[annot, midway, right=1.5mm] {scores attach to items, then\\ ride along every mention};
\node[stepnum] at ($(k3.north west)+(0.05,0)$) {B1};
\node[stepnum] at ($(k4.north west)+(0.05,0)$) {B2};
\node[stepnum] at ($(k5.north west)+(0.05,0)$) {B3};
\node[stepnum] at ($(k1.north west)+(0.05,0)$) {A};
\end{tikzpicture}}
\caption{Scoring and backfill, two tracks over one catalog. Track~A scores every item on the dimensions defined for its side with a grader selected by examination; track~B compiles per-bucket best-of-five regular expressions and streams them over the deduplicated descriptions with Hyperscan. Both tracks operate on the catalog rather than on the postings, which is what makes re-scoring and re-scanning bounded, repeatable operations. Colors as in Figure~\ref{fig:pipe-occ}.}
\label{fig:pipe-backfill}
\end{figure}
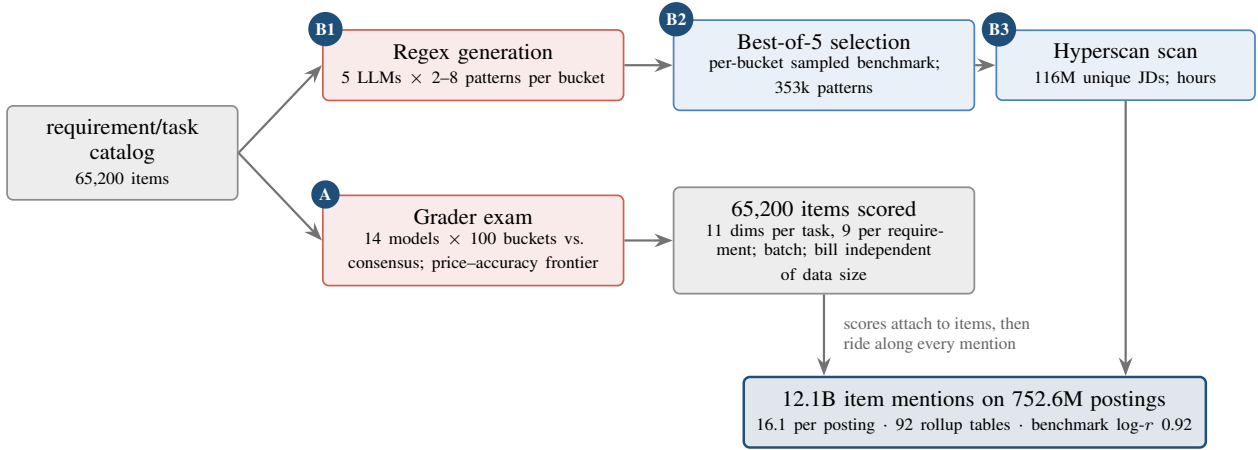

The catalog becomes data only when projected onto all 752.6~million postings (Figure~\ref{fig:pipe-backfill}), and this is where the introduction's third move, validation, lives: every pattern that will do the counting is benchmarked before it counts anything. For each bucket, five different LLMs independently generate two to eight Chinese regular expressions under tight constraints (they must anchor on core nouns and verbs, avoid matching generic filler, favor precision over recall, and remain compilable by Hyperscan, the high-throughput regular-expression engine used for the scan, which does not support backreferences or lookbehind assertions). Each model's pattern set is evaluated per bucket against a sampled LLM-annotated benchmark, and the best-performing model's patterns are retained \emph{for that bucket}; no model wins universally, and the per bucket selection outperforms any fixed single model. Regex generation and evaluation ran on all 44{,}690 task candidates that survived merging (65{,}411 candidates with the requirement side). Selection also closes the catalog: 21 candidate buckets whose best patterns still over-match and 190 that match nothing in the sample are dropped, fixing the final task catalog at 44{,}479; the scoring batch of Section~\ref{sec:scores} ran afterwards on the fixed 65{,}200-item master, so rejected candidates were never scored. The resulting 353{,}359 patterns across the full catalog (about 5.4 per bucket) are compiled into a Hyperscan database (a final compatibility pass rewrote 489 patterns in 345 buckets so that every pattern compiles) and streamed over the 116~million unique job descriptions; matches are joined back to the posting file. A full scan of the data completes in hours on a single workstation, which is what makes taxonomy evolution affordable: when the catalog changes, history is re-scanned rather than patched.

The backfill yields 12.1~billion bucket mentions, 16.1 per posting on average, each carrying the posting's month, city, occupation, industry, wage band, education and experience requirements, and firm size. The counting rule is bucket-level presence: a mention is a (bucket, posting) pair. Within a bucket, any number of matching spans in the description counts as one mention; across buckets there is no span-level deduplication: if one span satisfies patterns of two sibling buckets, each bucket records its own mention, so buckets act as independent binary detectors. Match sets are computed once per unique description (match positions are discarded) and inherited by every posting of that description. Accuracy against the extraction benchmark is quantified in Section~\ref{sec:validation}. For analysis convenience the mention file is pre-aggregated into 92 rollup tables (partitioned by platform, so that platform-subset series remain computable) covering all one-, two-, and three-way combinations of month, occupation, city, industry, wage band, education, experience, and firm size, each carrying bucket-category counts and the score sums needed to compute any weighted mean on the fly. All statistics and figures in Section~\ref{sec:facts} are computed from these rollups.

\section{Validation}
\label{sec:validation}

Every key stage of the construction is checked against a recorded reference before its output is trusted downstream, but the references differ in kind: some are external answer keys, some are the contracts that produced the stage, and the subsections below are ordered by that independence. Field standardization is judged against hand-labeled titles. Extraction and catalog quality are judged against the contracts that produced them, so what they establish is fidelity to a recorded rule rather than semantic truth. Backfill accuracy compares one machine output with another, the patterns against the extraction they are meant to reproduce. Score validity adds the first genuinely external test, agreement with published exposure measures built in other countries from other data (Table~\ref{tab:convergent}). Coverage, finally, is not an accuracy metric at all but a health check on the record as it refreshes. Table~\ref{tab:validation} consolidates the evidence stage by stage.

\begin{table}[t]
\centering
\caption{Validation summary, by construction stage.}
\label{tab:validation}
\footnotesize
\begin{tabularx}{\textwidth}{p{0.21\textwidth}p{0.26\textwidth}p{0.24\textwidth}X}
\toprule
\textbf{Stage} & \textbf{Sample / frame} & \textbf{Benchmark} & \textbf{Result} \\
\midrule
Occupation designs (\S\ref{sec:occ}) & Labeled title sample & Hand-labeled codes & Naive designs rejected (fabricated codes, 80--100\% error); retrieval-restricted selection error-free on sample; production design reproduces it via cached categories \\
Boundary adjudication (\S\ref{sec:reqres}) & $\sim$520{,}000 recalled pairs & Three-way contract, spot-checked & Merged catalog; disagreements resolved conservatively (\textsc{no} on doubt) \\
Regex backfill (\S\ref{sec:backfill}) & 44{,}690 task buckets $\times$ 5 models & Sampled extraction benchmark & Best-of-5 per bucket; modal model wins 32\% of buckets; 211 failing buckets dropped \\
Benchmark agreement (\S\ref{sec:backfill}) & about 44{,}500 buckets, nonzero benchmark & Extraction-sample frequencies & log-scale $r$ = 0.84 (0.92 freq.-weighted); median $|\Delta\log_{10}|$ = 0.14 \\
Grader exam (\S\ref{sec:scores}) & 100 buckets $\times$ 14 models & Consensus scores & MAE (mean absolute error) 0.061--0.124 across models; deployed grader 0.064, second-best, at four-fifths the exam cost of the most accurate \\
External convergence (\S\ref{sec:scores}) & 359--432 O*NET-SOC occupations & Published exposure measures & $r$ = 0.73--0.79 (Table~\ref{tab:convergent}); on-site vs.\ teleworkable $r=-0.63$ \\
\bottomrule
\end{tabularx}
\end{table}

\subsection{Field standardization}
This is the one stage judged directly against human labels. The four candidate occupation designs of Section~\ref{sec:occ} were benchmarked on a common hand-labeled title sample; the failure modes reported there (industry-vocabulary capture for embedding nearest neighbor, fabricated codes for closed-book LLM labeling) are measured on that sample, and the retrieval-restricted design whose judgment protocol the production crosswalk reuses made no errors on it. Because production resolves titles through cached platform-category judgments rather than per-title inference, its error concentrates where a platform category is wrong or missing: the embedding-vote path.

\subsection{Extraction and catalog quality}
Extraction and catalog quality are governed by contract rather than by an external answer key, so what follows establishes fidelity to those contracts. Extraction quality is governed by: atomic phrases only, meaning-preserving normalization, hard exclusions for benefits, boilerplate, and structured fields. Catalog quality is governed by the boundary-adjudication protocol, whose \textsc{same}/\textsc{loose}/\textsc{no} contract instructs the adjudicator to decide on core nouns and verbs and to return \textsc{no} on doubt, so that each judged pair errs toward over-segmentation, a direction conservative pair by pair: over-segmented buckets can be merged downstream, while contaminated buckets cannot be split without re-adjudication.

\subsection{Backfill accuracy}
Backfill accuracy is a machine-to-machine comparison: it asks whether the deployed patterns reproduce the extraction they were built from, measured against the extraction benchmark. on the 2.31-million-JD stratified sample, a bucket's benchmark frequency is the share of sampled descriptions whose LLM-extracted phrases map to it, and each candidate regex set is scored against that frequency during best-of-five selection, so every deployed pattern has a recorded benchmark performance. Across the roughly 44{,}500 task buckets with nonzero benchmark frequency, backfilled and benchmark frequencies agree at a log-scale correlation of 0.84 bucket-by-bucket, rising to 0.92 when buckets are weighted by benchmark frequency; the median absolute $\log_{10}$ gap is 0.14 (interquartile range 0.06--0.30). Because selection and evaluation share the same sampled benchmark, these agreement statistics are in-sample and should be read as optimistic, most of all for low-frequency buckets where a handful of sampled positives drive both the choice of pattern set and its evaluation. A held-out check now bounds that optimism on the task side: on a description-level 80/20 split of the benchmark sample, patterns re-selected on the training split alone agree with the held-out extraction at a log-scale correlation of 0.80 (0.91 frequency-weighted, median $|\Delta\log_{10}|$ of 0.165), against 0.84 in sample, and the published full-sample selection scores 0.81 on the same held-out split, so selection leakage is worth 0.011; the low-frequency caveat survives out of sample. On the requirement side the same split degrades less: patterns re-selected on the training split alone agree with the held-out extraction at 0.860 (0.982 frequency-weighted) against 0.871 in sample, the published full-sample selection scores 0.866 on the same held-out split, so selection leakage is worth 0.006, and the training-split re-selection picks the same generator as the full sample for 93.6\% of buckets. Selection is also a filter, and it is what closes the task catalog at 44{,}479 (Section~\ref{sec:backfill}). Precision is deliberately favored over recall in pattern construction (Section~\ref{sec:backfill}), so measured mention counts are best read as conservative. Full-catalog agreement on the requirement side is tighter than the task side on all three statistics: across the 20{,}818 evaluated candidates, backfilled and benchmark frequencies agree at 0.871 in logs, 0.983 when buckets are weighted by benchmark frequency, with a median absolute $\log_{10}$ gap of 0.125; selection closes the requirement catalog at 20{,}721, dropping 97 candidates, parallel to the task side's 211.

\subsection{Score validity}
Score validity is tested twice, once inside the model ensemble and once against the outside world. The grader exam scored fourteen candidate models on a common set of 100 buckets against a consensus benchmark; mean absolute errors ranged from 0.061 (most accurate, most expensive) to 0.124, and the deployed grader was chosen on the price--accuracy frontier at MAE~0.064. Because the deployed grader is itself a member of the fourteen-model consensus, exam error measures proximity to the model ensemble rather than to human judgment. Inter-model reliability is high on every dimension: single-model intraclass correlation ICC(2,1) ranges from 0.57 (human complementarity, the least reliably scored dimension) to 0.92 (embodied manual skill), with 0.81 for LLM exposure, and the consensus of all fourteen models reaches ICC(2,$k$) of 0.95--0.99, so the exam's benchmark is itself stable. Two further properties are worth recording as internal consistency rather than proof of validity: score distributions have intuitive frequency-weighted anchors on every dimension (Section~\ref{sec:scores}), and the two dimensions that ought to measure different things, LLM exposure and human complementarity, are indeed nearly uncorrelated ($r=-0.11$, Table~\ref{tab:scores}): handling customer complaints scores high on both. External convergent validity is assessed by aggregating our LLM-exposure dimension to O*NET-SOC occupations through the crosswalk of Section~\ref{sec:occ} (each occupation's score is a baseline-posting-weighted mean, a custom occupation's weight split equally across its SOC links, all joins at the six-digit SOC level) and correlating it with published exposure measures (Table~\ref{tab:convergent}). The alignment is strong: 0.73--0.77 with the GPT-exposure ratings ($\zeta$, the E1$+$E2 measure) of \citet{eloundou2024} (human and model labels respectively), 0.79 with the AI Occupational Exposure index, and 0.75 with its language-modeling variant \citep{felten2021}; furthermore, because both are high for desk and cognitive work, the correlation with teleworkability \citep{dingel2020} is 0.54. These correlations hold across 359--432 six-digit occupations, between measures built from different countries, languages, data sources, and methods; volume-weighted versions are higher still (up to 0.89 against AIOE and 0.83 against the model-label $\zeta$). The per-occupation inputs behind the table are committed at \texttt{figures/occ\_llm\_exposure.csv} (one row per six-digit SOC with the baseline-posting-weighted LLM-exposure and on-site-presence scores and the posting weight), so the table reproduces end to end from the external-validity package's script and benchmark file.

\begin{table}[t]
\centering
\caption{External convergent validity: correlation of our occupation-level LLM-exposure score with published AI/LLM exposure measures, via the O*NET-SOC crosswalk at the six-digit level (baseline-posting-weighted aggregation). All are positive, as expected.}\label{tab:convergent}
\footnotesize
\begin{tabularx}{\textwidth}{Xccc}
\toprule
\textbf{External measure} & \textbf{Pearson $r$} & \textbf{Spearman $\rho$} & \textbf{$n$} \\
\midrule
GPT exposure $\zeta$, human labels \citep{eloundou2024} & 0.73 & 0.74 & 432 \\
GPT exposure $\zeta$, model labels \citep{eloundou2024} & 0.77 & 0.77 & 432 \\
AI Occupational Exposure \citep{felten2021}             & 0.79 & 0.77 & 359 \\
Language-Modeling AIOE \citep{felten2021}               & 0.75 & 0.73 & 359 \\
Teleworkability \citep{dingel2020}                      & 0.54 & 0.56 & 359 \\
\bottomrule
\end{tabularx}
\end{table}

As a discriminant check, our on-site-presence dimension correlates \emph{negatively} with teleworkability ($r=-0.63$, $\rho=-0.64$; volume-weighted $-0.74$), the opposite sign of LLM exposure, and it is likewise negative against every convergent measure ($-0.69$ to $-0.74$ against the \citealp{eloundou2024} and \citealp{felten2021} indices). The two dimensions measure different things, and each aligns with the external measures in the expected direction.

\subsection{Coverage}
The backfill attaches 12.1~billion mentions to 752.6~million postings, 16.1 per posting on average, split almost evenly between requirements (6.13~billion) and tasks (5.98~billion), so neither object dominates the record. Two health metrics are computed with each monthly refresh: the share of postings with no catalog match, and the share of newly extracted phrases that match no existing bucket. The latter doubles as a new-work detector in the spirit of \citet{autor2024newfrontiers}: a sustained rise flags vocabulary the market has invented since the catalog was last derived, and triggers re-derivation. On the current record, fewer than 0.005\% of postings match nothing on either side. The one-sided rates are naturally higher and are monitored directly, because time-varying recall would be a first-order threat to the time-difference results of Section~\ref{sec:facts}: the share of postings with no \emph{task}-side match sits near 7\% across the analysis window (7.1\% in 2023Q4 against 7.5\% in January--April 2026 on the five-platform pool; 7.0\% against 7.2\% on the stable-coverage subset), with no sustained deterioration, and, because the catalog was derived in early 2026 from the full 2022--2026 data, no scope for staleness at the window's end. Within every platform separately, the share of postings mentioning at least one top-quartile-exposure item is essentially flat over the window (within four percentage points on each platform) while low-exposure presence rises on all five, so the aggregate exposure decline of Section~\ref{sec:portraits} reflects expanding low-exposure content and shifting composition rather than a collapse of high-exposure matching; these are posting-level indicators, so they bound vocabulary coverage rather than per-mention recall, and the held-out check above, cut by posting year, finds the correlation flat at 0.71--0.73 from 2022 through 2026, so per-mention match quality shows no drift either; the monthly series are released at \texttt{issue1\_results/}. The mentions-per-posting distribution has a 10th percentile of 3, a median of 14, and a 90th percentile of 32: nearly every ad, not just the verbose minority, contributes task-level signal. The college-major field is sparser by construction, since it exists only where employers restrict majors: 28.3\% of postings carry a major requirement matched to a Ministry of Education catalog entry (undergraduate or vocational system).
 Taken together, these checks establish reproducibility relative to the recorded contracts, not accuracy against human judgment: the deliberate limit of a program benchmarked end to end by machine.
\section{What the record can show}
\label{sec:facts}

Two examples follow, and they do different jobs. The first shows what the record can measure: which job titles hold together as descriptions of work and which do not, charted title by title, across market segments and over time. The second shows why that measurement matters, by going one layer beneath a pattern other instruments have begun to document: where posted demand for the work most exposed to language models is contracting, and where it is not. Throughout, six large occupations serve as running cases (accountant, Java developer, customer-service representative, sales consultant, production worker, and food-delivery rider), chosen to span the digital/hands-on and cognitive/interactive spectrum. Platform pools differ by analysis and each figure states its own: the drift series of this section use the three platforms with the longest consistent coverage (A, B, D), the two-layer exercise of Section~\ref{sec:twolayers} uses the two largest platforms (A, B), and composition snapshots use the full five-platform pool.

\subsection{Example 1: the staircase inside a title}
\label{sec:portraits}

\begin{figure}[t]
\centering
\includegraphics[width=\textwidth]{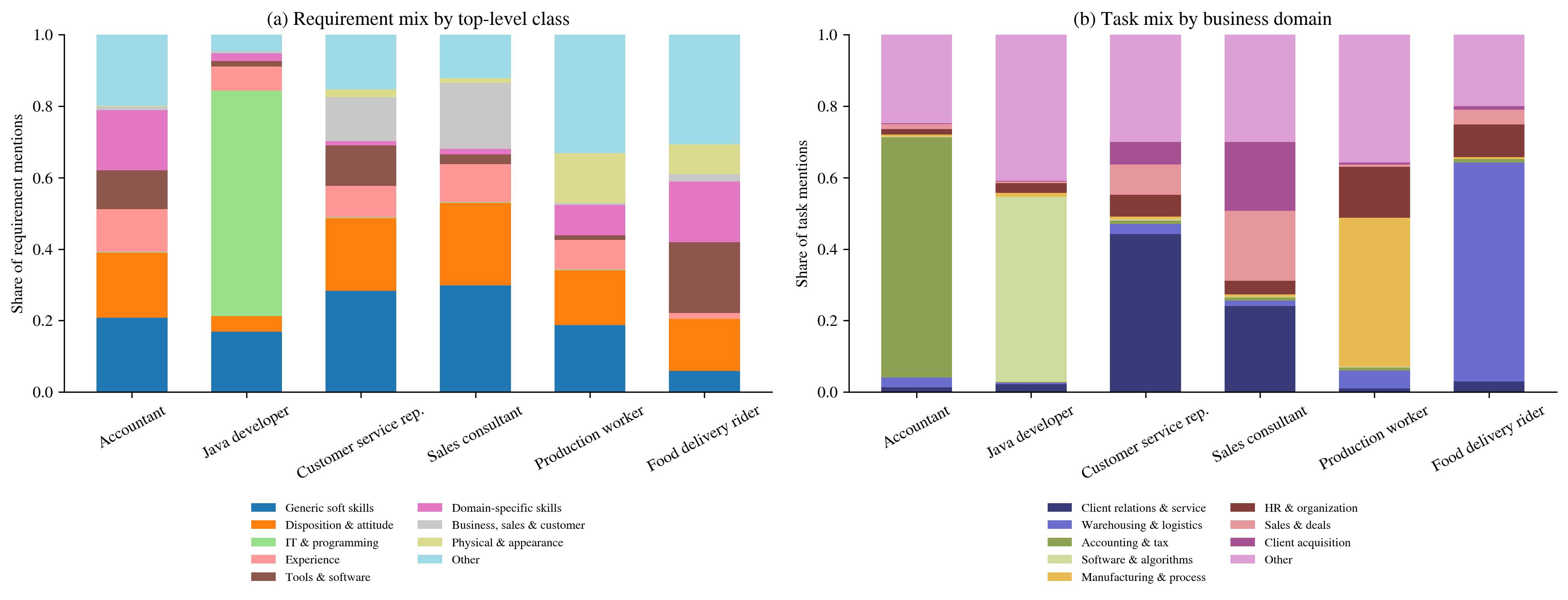}
\caption{Task portraits of six occupations. Each stacked bar decomposes one occupation's mentions (matches of catalog items in the text of its job ads) into shares that sum to one. Panel (a) decomposes \emph{requirement} mentions (what the candidate must satisfy: credentials, skills, disposition) across the top-level classes of the requirement taxonomy; panel (b) decomposes \emph{task} mentions (what the hire will actually do) across business domains (the eight largest shown; the remainder grouped as Other). Sample: all postings in these occupations on the five platforms, January--April 2026. Both panels show the eight largest categories, with the remainder grouped as Other. Task portraits are sharp (accountants concentrate in accounting \& tax, Java developers in software \& algorithms, delivery riders in warehousing \& logistics, with no supervision imposed), while requirement portraits carry a thick band of generic soft skills and disposition in the service and clerical occupations, with specialist bands dominating for the developer (IT \& programming) and the rider (tools and physical items). The two sides of an ad therefore measure different things, which is why the dataset keeps them as separate objects.}
\label{fig:composition}
\end{figure}

A job title carries real information: title words alone predict pay and applications \citep{marinescu2020}. What no title-level dataset can say is whether two postings sharing a title describe the same work. The task layer turns that question into a measurement, and the first thing to check is that the instrument reads what it claims to read. Figure~\ref{fig:composition} shows each occupation's requirement mix (by top-level class) and task mix (by business domain) for January--April 2026. Three features validate the instrument. First, task portraits are sharp: accountants' tasks concentrate in accounting \& tax, Java developers' in software \& algorithms, riders' in warehousing \& logistics. No supervision forced this; it emerges from employer text. Second, requirement portraits are \emph{softer} than task portraits for most titles: the service and clerical occupations demand a thick band of generic soft skills and disposition, consistent with \citet{deming2017}. The delivery rider and the Java developer are the exceptions, dominated instead by their specialist bands; what differentiates occupations on the requirement side is the specialist band (IT \& programming for developers, licenses and domain skills for accountants, physical and work-context requirements for hands-on work). Third, the two sides do not mirror each other: customer-service representatives' requirements emphasize disposition while their tasks are dominated by client relations \emph{operations}. That is why the system keeps requirements and tasks as separate objects.

\begin{figure}[t]
\centering
\includegraphics[width=\textwidth]{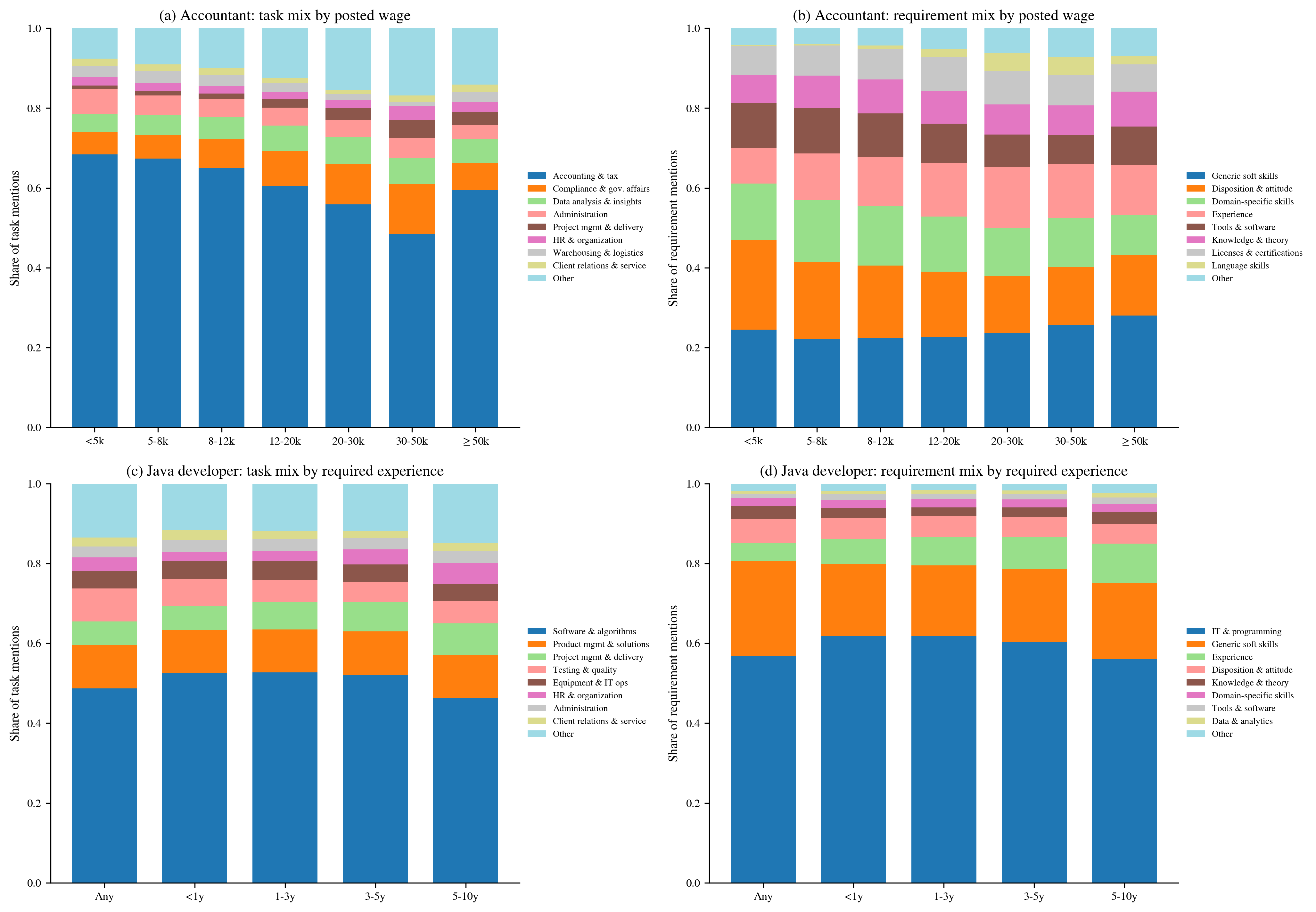}
\caption{Both sides of one job title, cut within the title. The same objects as Figure~\ref{fig:composition}, task mentions by business domain and requirement mentions by top-level class, but split \emph{inside} a single title instead of across occupations. Panels (a) and (b): accountant postings split by posted monthly wage (thousand CNY per month). Panels (c) and (d): Java-developer postings split by required experience. Sample: five platforms, January 2025--April 2026. As the posted accountant wage rises from under 5k to 30--50k, the accounting-and-tax core falls from 68\% to 49\% of task mentions while compliance \& government affairs, data analysis, and administration expand, and on the requirement side disposition and attitude give way to experience, language, and credentials. The licences class itself barely moves; the certificates inside it change rank (Table~\ref{tab:ladder}). Java postings carry essentially the same software-and-algorithms core at every experience level and an equally stable requirement mix, so for some occupations the occupation average is a good sufficient statistic on both sides at once. The 50k-and-above band is not read: on one platform it carries salary values recorded in annual units, which places mid-level postings in the top cell.}
\label{fig:within}
\end{figure}

Within a single title the variation is just as sharp, and it runs on both sides at once. Figure~\ref{fig:within} cuts two occupations four ways. Panel~(a): as the posted wage of \emph{accountant} rises from under \yen5k to \yen30--50k per month, the share of core bookkeeping tasks (accounting \& tax) falls from 68\% to 49\%, while compliance \& government affairs, data analysis, and administration expand. The title is the same; the job is not.

Panel~(b) asks what it takes to get in. Read as a share of requirement text, the answer looks like nothing at all: the licences-and-certifications class moves only from 7.2\% to 8.5\% across the same range, holding a near-constant slice of requirement lists that lengthen with pay. Read directly, as the share of postings demanding at least one certificate, the bar is anything but flat: 28\% of accountant postings below \yen5k, 52\% at \yen12--20k, and there it stops, with the two bands above holding at 51\%. Below the plateau employers ask for more certificates; on the plateau the asking changes character, and the item level shows how. Table~\ref{tab:ladder} shows the four certificates accountants are actually asked for. The entry-level title peaks early and then withdraws, from 14.5\% of postings at \yen5--8k to 4.7\% at \yen20--30k, while the intermediate title rises from 1.9\% of postings to a peak of 14.1\% at \yen12--20k, the certified-public-accountant qualification more than threefold, and the tax-accountant licence nearly fourfold, each peaking in the same band. Past \yen12--20k employers are not asking for more certificates. They are asking for a different one. A sixteen-class taxonomy averages both movements into a straight line; recovering them is what the 20{,}721-entry vocabulary is for.

The rest of the requirement side moves with the same logic. Across the same range the disposition-and-attitude class falls from 22.4\% of requirement mentions to 14.2\%, while experience rises from 8.8\% to 15.2\% and language skills from 0.4\% to 4.4\%. What the bottom of the range spends its requirement text on is character, carefulness and responsibility and diligence, assertions a candidate can make at no cost and an employer cannot check. What the top spends it on is verifiable: years on the job, a language certificate, a professional title. Where a credential a third party administers can separate candidates, employers stop asking candidates to vouch for themselves. Shares sum to one, so the identity guarantees that something falls when the verifiable band rises; it does not guarantee that disposition is what falls. Disposition is the largest single faller in the table, at $-8.1$ points against $-3.1$ for tools and software and $-2.2$ for domain-specific skills.

Panels~(c) and~(d) supply the counterexample, and it holds on both sides. Java developer postings carry nearly the same software-and-algorithms core at every experience level, and their requirement mix moves as little. Between the lowest and highest cells the Jensen--Shannon divergence of the accountant's task mix is 0.037 and of its requirement mix 0.034; for the Java developer the two numbers are 0.013 and 0.011. Both sides of the accountant's title move about three times as much as the Java developer's, and neither side of the Java developer's title moves much at all. Whether an occupation's average is a good summary is therefore a property of the occupation rather than of the side one happens to read, and knowing \emph{which} occupations average well is itself something only within-occupation data can tell you.

Two cells are left out of these readings. The \yen50k-and-above band carries salary values recorded in annual rather than monthly units on one platform, where the median posted figure reaches \yen290{,}000 and two postings in five match no requirement at all; its apparent rebound in the accounting-and-tax core is an artifact of that recording, not a feature of the market. The item-level table is read to \yen20--30k, the last cell holding more than thirty thousand postings.

\begin{table}[t]
\centering
\caption{The certificate staircase inside one job title. Share of \emph{accountant} postings requiring each certificate, by posted monthly wage. The licences-and-certifications class holds a near-constant share of requirement mentions across these cells (7.2\% to 8.5\%, Figure~\ref{fig:within}b) while the share of postings demanding at least one certificate nearly doubles and then saturates; the further movement is inside the class. Sample: five platforms, January 2025--April 2026. Item and incidence shares are computed over postings, the class shares of Figure~\ref{fig:within} over mentions.}
\label{tab:ladder}
\footnotesize
\begin{tabular}{lrrrrr}
\toprule
\textbf{Certificate} & \textbf{$<$\yen5k} & \textbf{\yen5--8k} & \textbf{\yen8--12k} & \textbf{\yen12--20k} & \textbf{\yen20--30k} \\
\midrule
At least one certificate & 28.0 & 39.6 & 46.5 & \textbf{51.9} & 51.5 \\
\midrule
Junior accountant title & 12.5 & \textbf{14.5} & 12.3 & 7.1 & 4.7 \\
Intermediate accountant title & 1.9 & 6.4 & 12.0 & \textbf{14.1} & 9.2 \\
Certified public accountant & 1.5 & 2.7 & 3.9 & \textbf{5.3} & 4.0 \\
Tax accountant licence & 1.2 & 2.3 & 3.7 & \textbf{4.6} & 4.0 \\
\midrule
Postings in cell (thousands) & 992 & 2{,}129 & 836 & 204 & 33 \\
\bottomrule
\end{tabular}
\end{table}

\begin{figure}[t]
\centering
\includegraphics[width=\textwidth]{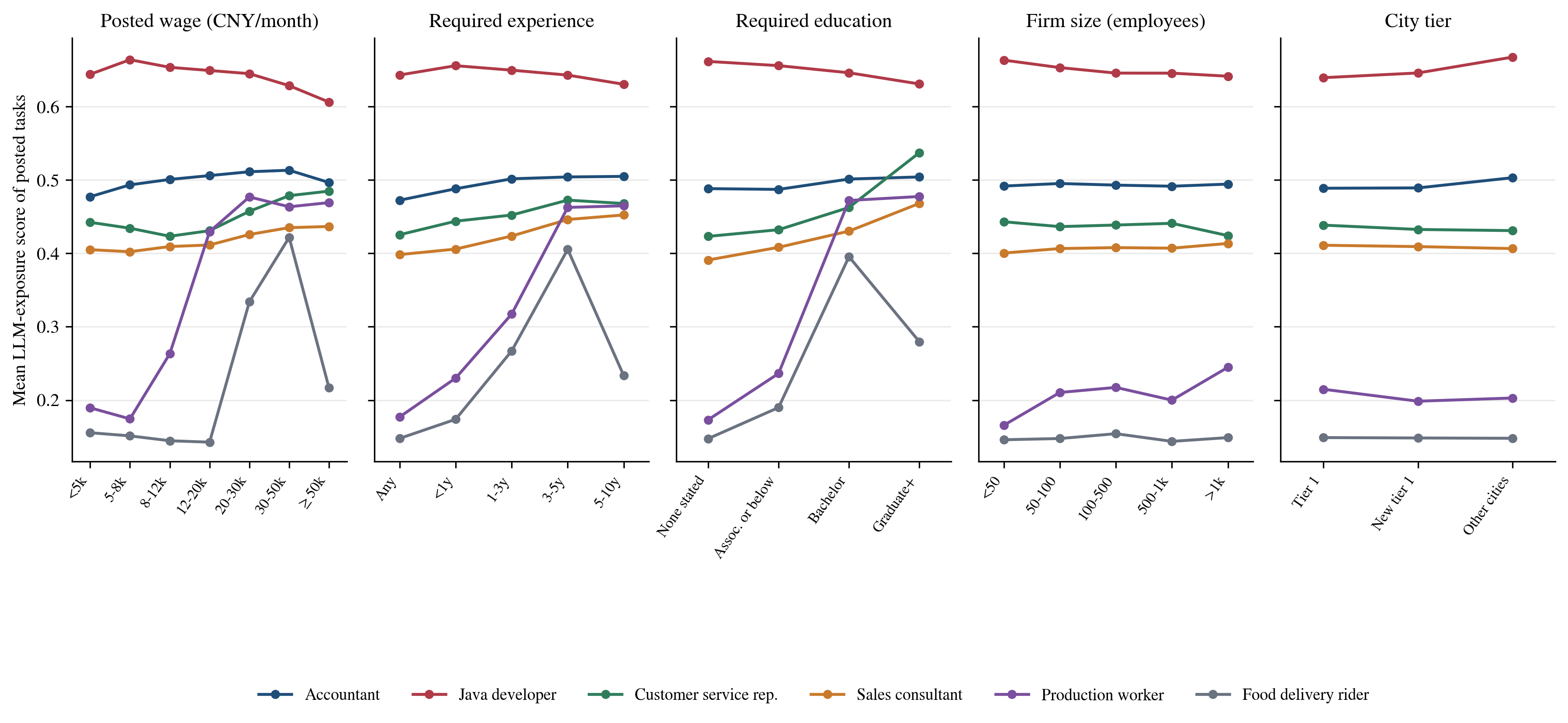}
\caption{Mean LLM-exposure score of posted tasks within occupation, by market segment. Each point is the mention-weighted mean of the 0--1 LLM-exposure scores of all mentions in postings of one occupation within one segment cell (posted wage band, required experience, required education, employer size, or city tier). LLM exposure scores each catalog item on whether current large language models can substantially accelerate or absorb its core output (Section~\ref{sec:scores}). Sample: five platforms, January 2025--April 2026. The white-collar titles are comparatively flat along every dimension (an accountant's task mix carries the same exposure in Tier-1 cities as elsewhere), while blue-collar titles are steep along the wage and skill margins: the mean exposure of production-worker postings rises from below 0.2 in the lowest wage bands to $\approx0.48$ at 20--50k CNY per month, where ads describe quality documentation and process coordination rather than assembly; delivery riders rise comparably steeply. The 50k-and-above cell is not read; on one platform it carries salary values recorded in annual units. Among the six occupations shown, occupation-level exposure scores are most misleading for the blue-collar titles.}
\label{fig:het}
\end{figure}

Figure~\ref{fig:het} generalizes the point to a score. For each occupation we compute the mean LLM exposure of its posted tasks within wage, experience, education, firm-size, and city-tier cells. Two regularities stand out. First, the white-collar titles are comparatively flat across the five dimensions: an accountant's task mix carries the same exposure in Tier-1 cities as elsewhere, in small firms as in large. Second, blue-collar titles are dramatically heterogeneous along the wage and skill margins: the mean exposure of \emph{production worker} postings rises from below 0.2 in the lowest wage bands to about 0.48 at \yen20--50k, where high-wage ``production worker'' ads describe quality documentation, data recording, and process coordination rather than assembly, and delivery riders rise just as steeply through the \yen30--50k band. The top cell is not read, for the recording reason given above. Among the six occupations tracked here, occupation-level exposure scores are most misleading for the blue-collar titles, production workers and riders, where policy attention concentrates. City-tier profiles, by contrast, are almost flat: task content within an occupation is national; what differs across places is the occupation mix.

\begin{figure}[t]
\centering
\includegraphics[width=\textwidth]{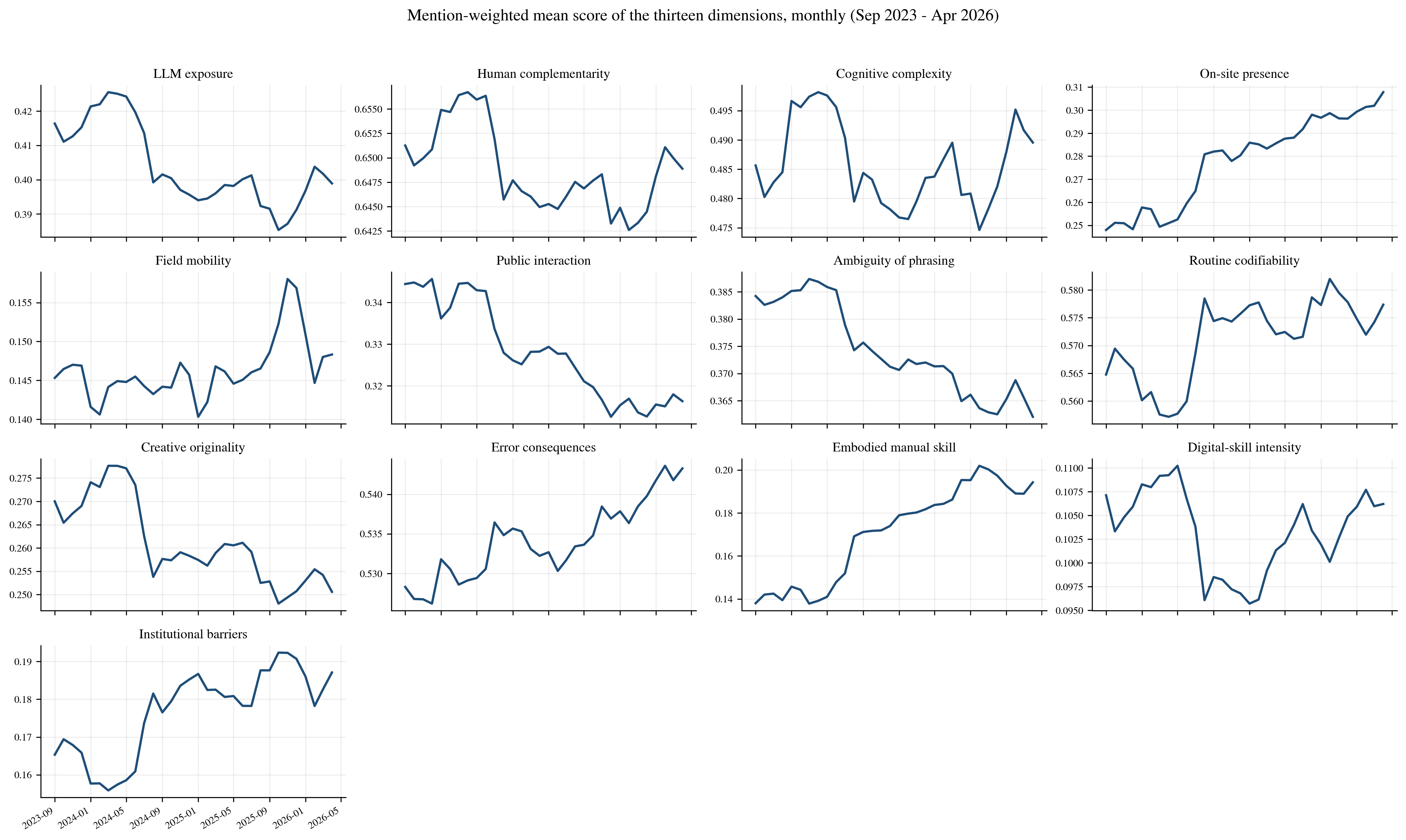}
\caption{Mention-weighted mean score of the thirteen task dimensions, monthly. Each panel tracks the mention-weighted mean of one 0--1 dimension score across all mentions on the three platforms with stable coverage (Platforms A, B, D), September 2023--April 2026. Panel y-axis ranges differ, and some span only a few hundredths, so slopes are not visually comparable across panels. Scores are assigned per catalog item and attach to every mention; requirement-only dimensions (digital-skill intensity, institutional barriers) average over requirement mentions, task-only dimensions over task mentions, and shared dimensions over both. LLM exposure declines from its mid-2024 level, alongside public interaction, phrasing ambiguity, and creative originality, while on-site presence, embodied manual skill, error consequences, and institutional barriers rise over the window; the series move in consistent directions but measure distinct constructs (Table~\ref{tab:scores}). Posted work on these platforms is, month by month, becoming more physical, more consequence-supervised, more credential-gated, and less like the text-and-analysis work that LLMs perform.}
\label{fig:dim13}
\end{figure}

\paragraph{The same chart, over time.}
A title-level series can count postings; it cannot say whether the \emph{content} of posted work is changing, nor, if it is, whether the change happens within occupations or between them. The task layer answers both. Figure~\ref{fig:dim13} tracks the mention-weighted mean of all thirteen dimensions monthly from September 2023 to April 2026 on the three platforms with the longest consistent coverage. The headline movement is direct: LLM exposure itself is declining. LLM exposure falls from a mid-2024 peak (a net change of about $-0.012$ on the 0--1 scale from the 2023Q4 baseline, roughly 3\% of its level), public interaction and phrasing ambiguity decline steadily, creative originality drifts down. Dimensions describing what models cannot absorb rise: on-site presence (+0.06 over the window), embodied manual skill (+0.06), error consequences, institutional barriers. These are related readings rather than one axis: Table~\ref{tab:scores} shows public interaction, error consequences, and institutional barriers nearly uncorrelated with LLM exposure across the catalog, so each series is its own fact. Read together, posted work on these platforms is, month by month, becoming more physical, more supervised-by-consequences, more credential-gated, and less like the text-and-analysis work that LLMs perform.

\begin{figure}[t]
\centering
\includegraphics[width=0.9\textwidth]{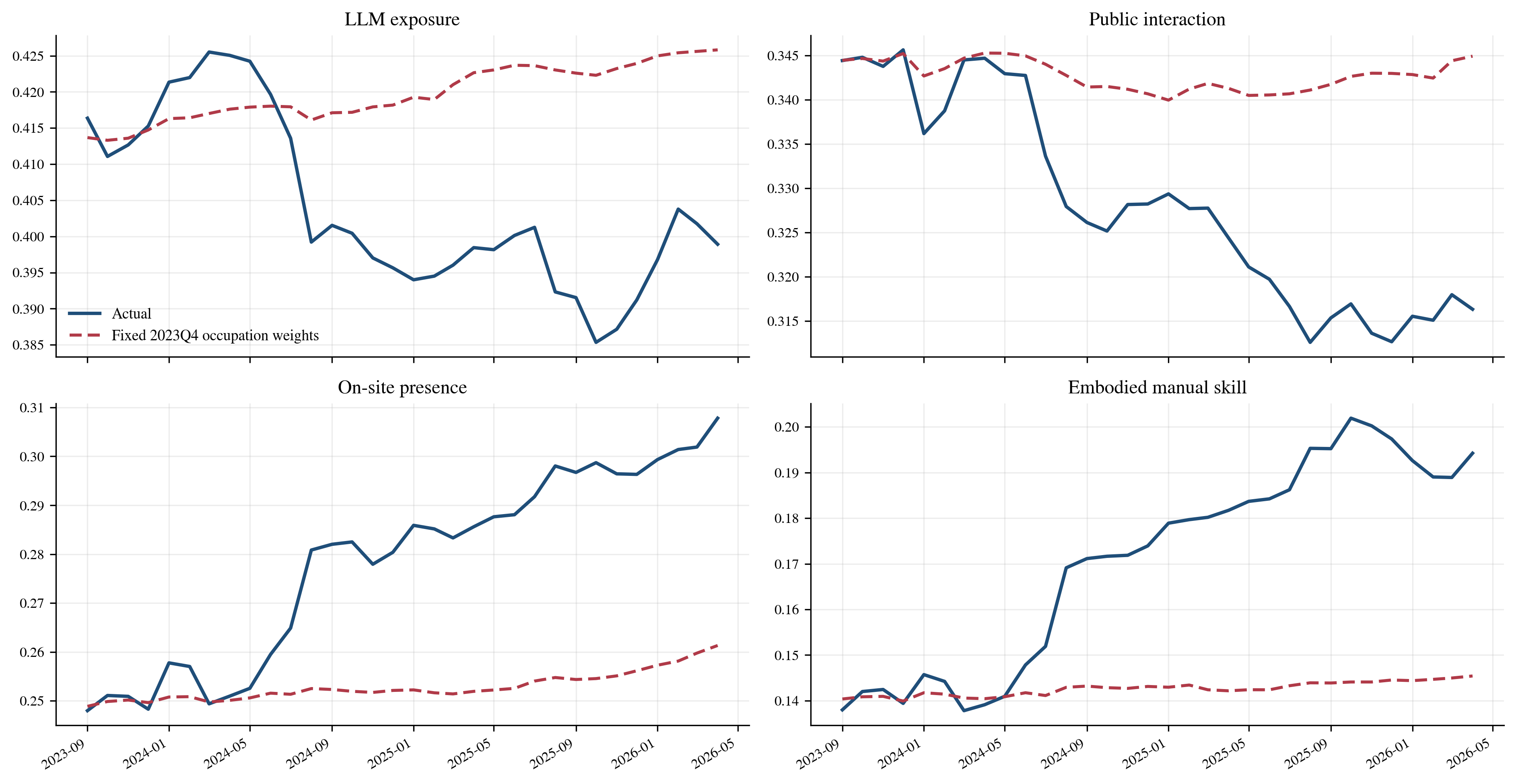}
\caption{Actual versus fixed-composition counterfactual task-dimension series. Solid lines: mention-weighted monthly mean of four dimension scores (0--1) across all mentions on the three platforms with stable coverage (Platforms A, B, D), September 2023--April 2026. Dashed lines: a counterfactual that freezes every occupation's \emph{mention} weight at its 2023Q4 value and averages occupation-level mean scores under those frozen weights, so that only within-occupation change remains. Both series use the same mention-weighting basis, so the counterfactual's base-quarter average coincides with the actual by construction (occupations entering after the base quarter carry zero weight; weights renormalize over the occupations present each month), and the vertical gap between the lines \emph{is} the contribution of occupation re-weighting. That gap is nearly the whole story: for LLM exposure the fixed-weight series drifts mildly \emph{upward} (+0.012) while the actual falls ($-$0.012); the fixed-weight lines capture about a fifth of the actual rise in on-site presence (+0.010 of +0.054) and less than a tenth in embodied manual skill (+0.004 of +0.050). The aggregate drift of posted work, away from what language models absorb and toward hands-on, on-site work, is therefore chiefly a re-weighting across occupations rather than a rewriting of any occupation's own task mix.}
\label{fig:counterfactual}
\end{figure}

\begin{figure}[t]
\centering
\includegraphics[width=0.9\textwidth]{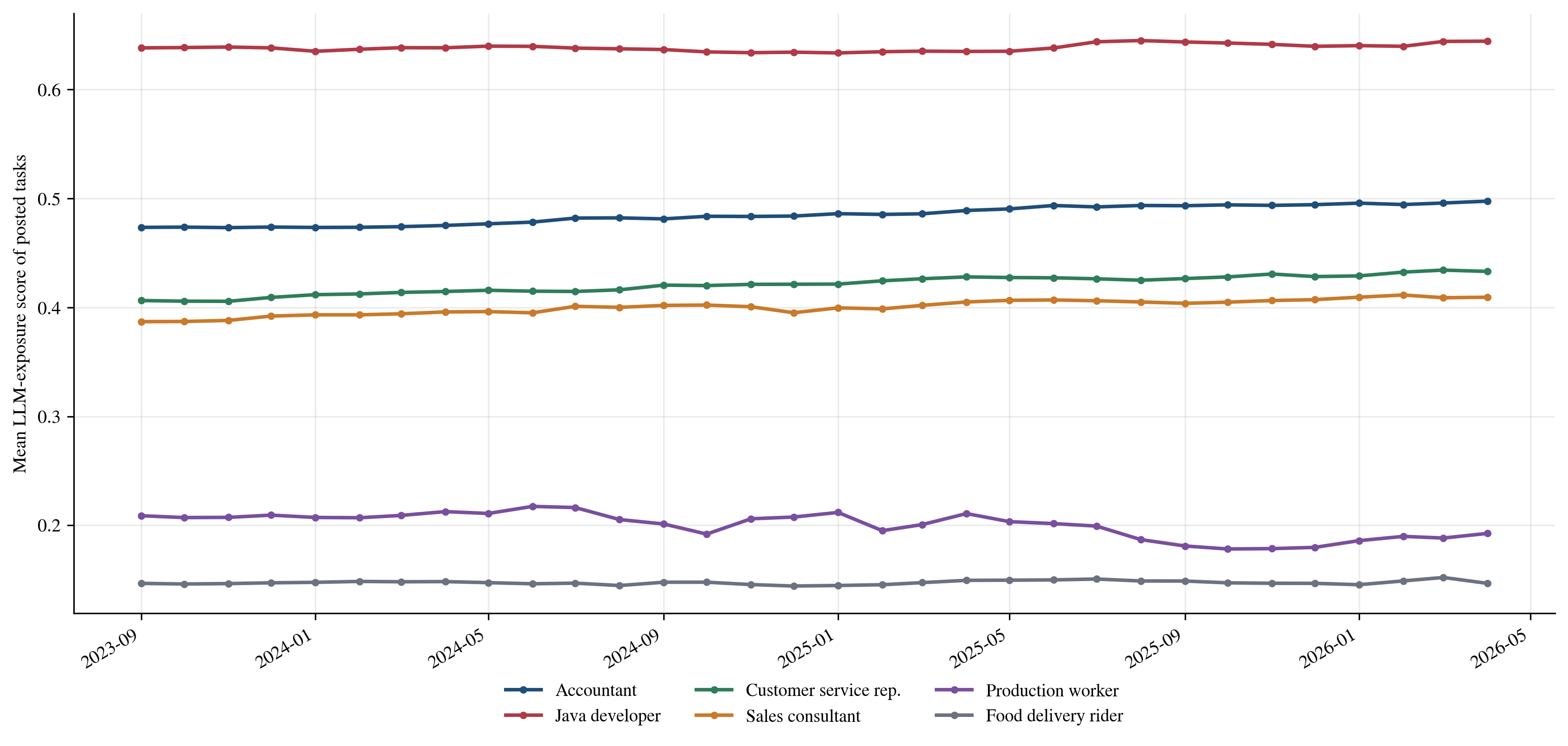}
\caption{Mean LLM-exposure score of posted tasks, monthly, for six occupations. Each line is the mention-weighted monthly mean LLM-exposure score (0--1) of one occupation's mentions, on the three platforms with stable coverage (Platforms A, B, D), September 2023--April 2026. Levels separate cleanly and persistently: Java developer $\approx0.64$, accountant $\approx0.49$, food-delivery rider $\approx0.15$, and every line moves within about $\pm0.03$ over 32 months. Occupations' own exposure levels are essentially fixed; the aggregate decline in posted LLM exposure (Figure~\ref{fig:dim13}) therefore travels through the changing occupation mix, not through any occupation's task content.}
\label{fig:occtrend}
\end{figure}

\begin{figure}[t]
\centering
\includegraphics[width=\textwidth]{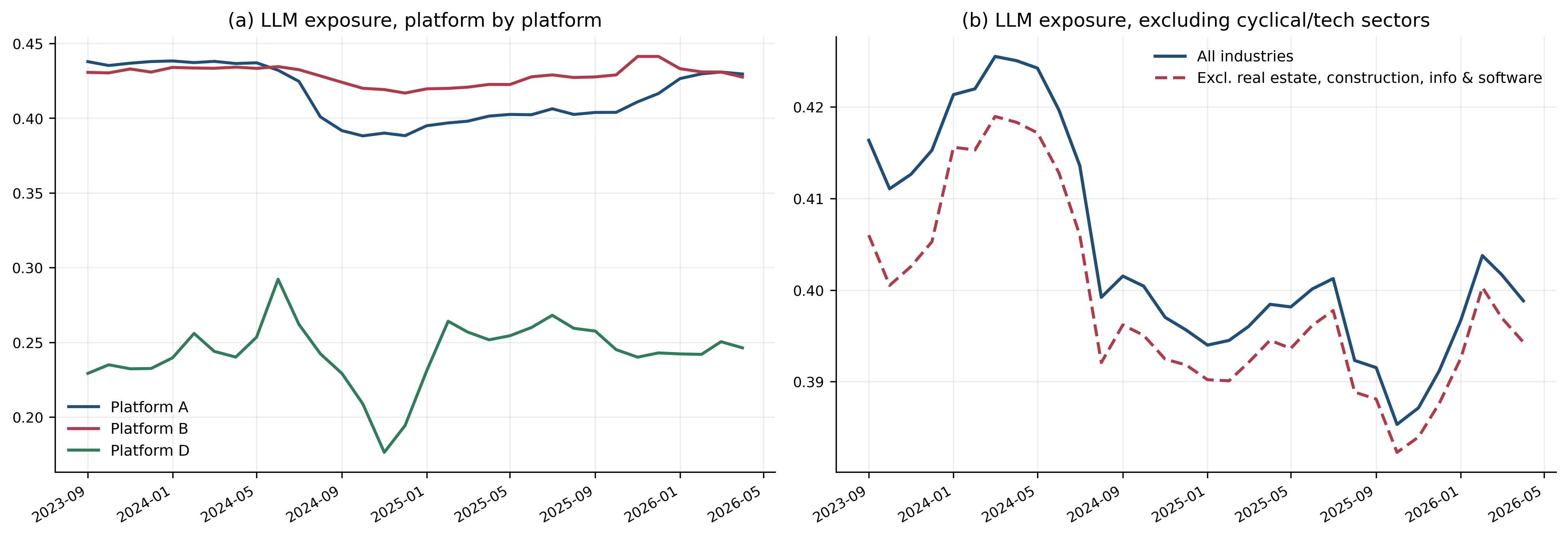}
\caption{Robustness of the post-2024 decline in posted LLM exposure. The quantity plotted is the mention-weighted monthly mean LLM-exposure score (0--1) across mentions, September 2023--April 2026. Panel (a) recomputes the series separately on each of the three stable-coverage platforms (A, B, D): levels differ with each platform's client mix, but the decline from the mid-2024 peak appears in each. Panel (b) recomputes the aggregate over all industries (solid) and after excluding real estate, construction, and information \& software (dashed), the three sectors whose own cycles (a property downturn, a tech-hiring correction) could most plausibly masquerade as task-content drift. The decline survives both cuts.}
\label{fig:driftrobust}
\end{figure}

\begin{figure}[t]
\centering
\includegraphics[width=\textwidth]{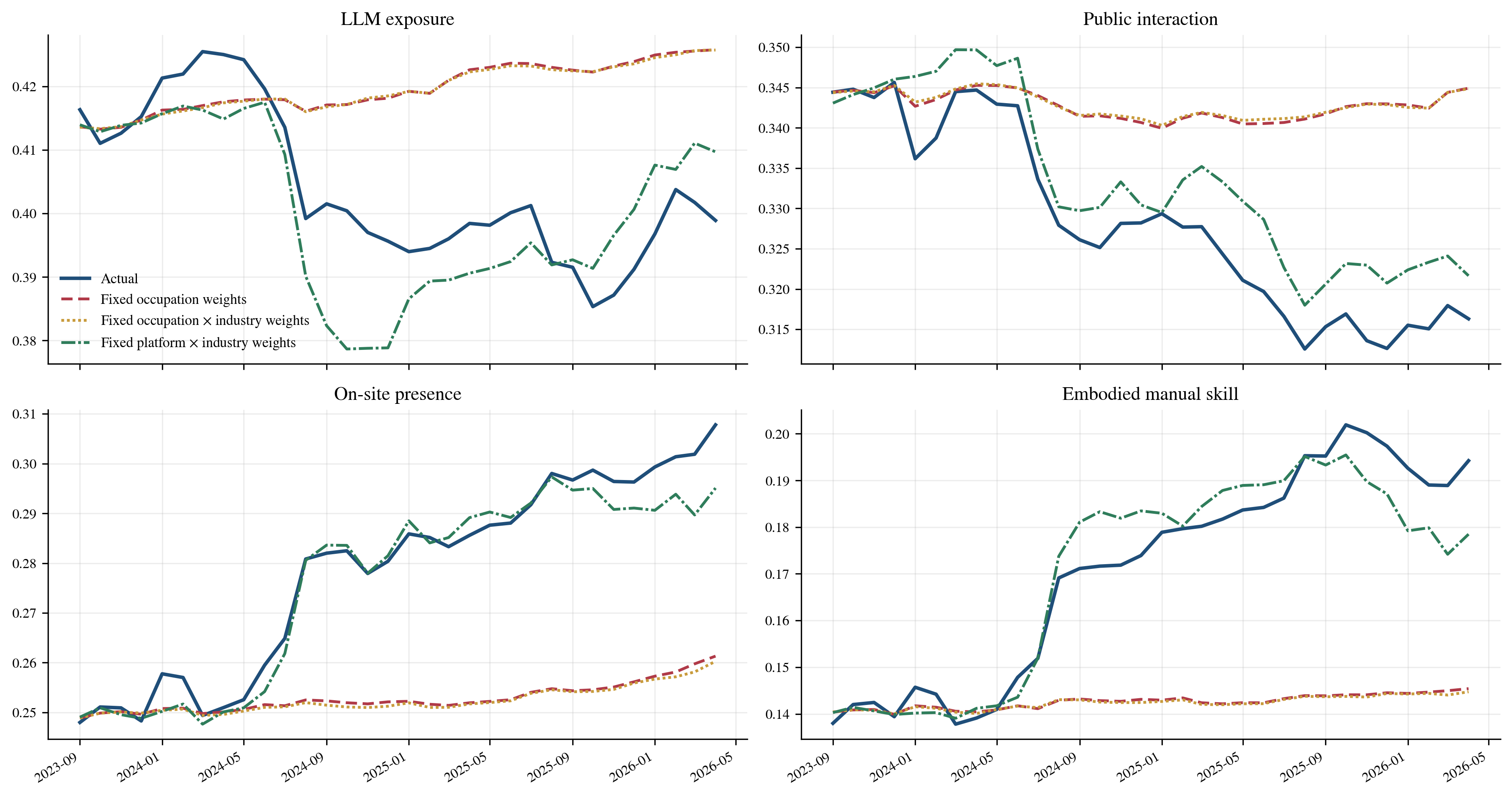}
\caption{Upgraded fixed-composition counterfactuals. Actual (solid) mention-weighted mean scores for four dimensions, monthly, September 2023--April 2026 (three stable-coverage platforms), against three counterfactual series: occupation mention weights frozen at 2023Q4 (as in Figure~\ref{fig:counterfactual}); occupation $\times$ industry cell mention weights frozen (22{,}338 cells); and platform $\times$ industry cell mention weights frozen. All series share the mention-weighting basis, so each counterfactual's base-quarter average coincides with the actual and vertical gaps read directly as composition contributions. The occupation-level and occupation $\times$ industry counterfactuals are nearly identical on all four dimensions: industry cycles operating inside occupations account for none of the within-occupation component; the platform $\times$ industry freeze recovers more of the on-site and embodied rise but reverses none of the reading.}
\label{fig:cfup}
\end{figure}

The decomposition is formalized in Figure~\ref{fig:counterfactual}, which replays each dimension holding occupation mention weights fixed at their 2023Q4 values, so that only within-occupation change remains; because actual and counterfactual share the same weighting basis, the counterfactual passes through the actual in the base quarter and the gap between the lines is the composition contribution itself. The counterfactual series are far flatter than the actual ones on every dimension shown: for LLM exposure the fixed-weight series drifts mildly \emph{upward} (+0.012 over the window) while the actual series falls ($-$0.012), and the fixed-weight series capture only about a fifth of the actual rise in on-site presence (+0.010 of +0.054) and less than a tenth of the rise in embodied manual skill (+0.004 of +0.050). Consistent with this, the six tracked occupations' own exposure levels move within about $\pm0.03$ over 32 months, with clean level separation (Java $\approx$0.64, accountant $\approx$0.49, rider $\approx$0.15; Figure~\ref{fig:occtrend}). Composition therefore sets the sign and the bulk of the aggregate drift: the posting pool is re-weighting away from high-exposure occupations faster than surviving occupations' own content moves. A drift of this shape is consistent both with technology reshaping task demand and with cyclical forces re-weighting sectors: the deepest contractions include sales-related work, which a demand-cycle story also predicts, so this section documents the anatomy of the change, not its cause. And the within-occupation component is material rather than decorative: on LLM exposure the fixed-weight series rises $+0.012$ while the actual series falls $-0.012$, so the implied composition pull of $-0.024$ is half offset from within, which is exactly the recomposition margin that Example~2 takes up. Figure~\ref{fig:driftrobust} probes the headline dimension both ways: the decline in LLM exposure is present within each of the three platforms separately, and it survives recomputing the aggregate over all industries except real estate, construction, and information \& software, the three sectors whose own cycles (a property downturn, a tech-hiring correction) could most plausibly masquerade as task-content drift. A tighter counterfactual reaches the same answer (Figure~\ref{fig:cfup}): holding occupation $\times$ industry cells fixed (22{,}338 cells) rather than occupations alone reproduces the occupation-level counterfactual almost exactly on all four dimensions: the within-occupation component is unchanged when industry cycles operating \emph{inside} occupations (a property-sector contraction among accountants) are explicitly frozen. Holding platform $\times$ industry cells fixed moves the series only modestly back toward the actual.

\subsection{Example 2: occupations contract faster than the tasks they contain}
\label{sec:twolayers}

In U.S.\ payroll data, employment has begun to fall in the occupations most exposed to AI, with early-career workers hit first \citep{brynjolfsson2025canaries}. Posted demand is a different object from employment, and China is a different market from the United States, but the record built here can ask what such a contraction looks like one layer down. Going down starts with a definition. In this system an occupation is exposed exactly to the degree that its tasks are: occupation-level exposure is the frequency-weighted mean of the exposure scores of the tasks its postings mention, so an occupation is nothing but a weighted average of its tasks. The distinction matters, because a task can sit in occupations of very different exposure, and its fate need not follow the fate of any one of them.

\paragraph{Design.} The second example uses the system's two native layers at once: the joint measurement of occupations and tasks in the same data that no single-layer dataset, title-based or task-based, offers. The exercise fits two related descriptive gradients, one at each layer. At the \emph{occupation layer}, each observation is an occupation: we compute the change in log posting share between the 2023Q4 baseline and January--April 2026 (the two largest platforms, A and B) and relate it to the occupation's baseline mean LLM exposure. The sample keeps occupations with at least 1{,}000 postings in \emph{both} the baseline and the endline window; occupations that exit, enter, or fall below the threshold in either window are excluded, so the estimates describe continuing occupations, and the truncation clips the most extreme contractions and expansions out of both tails. The clipping is conservative for the gradient, not the source of it: switching to an arc-style share change defined at zero and including all 1{,}267 occupations, entrants and exiters among them, steepens the slope to $-2.01$ volume-weighted ($-2.38$ unweighted), against $-1.24$ for the same arc measure on the filtered sample; the 282 occupations the threshold drops carry 0.33\% of baseline postings and are on average less exposed than those kept (0.37 vs.\ 0.43). At the \emph{task layer}, each observation is one occupied cell of the $32\times26$ task grid: the same change in log mention share, related to the cell's frequency-weighted mean exposure. Of the 698 occupied cells (Section~\ref{sec:reqres}), one never appears in the two-platform estimation sample, leaving $n=697$ throughout this section. The two gradients share the same exposure scale, the same window, and the same data, which makes their slopes comparable as descriptions. What they do not share is the outcome variable, its denominator, the weighting, or the level of aggregation, so the vertical distance between the fitted lines is a comparison of two descriptions, not an identified decomposition. The decomposition comes later, and separately.

\paragraph{What the slopes are not.} These are descriptive gradients, not causal estimates, and because exposure enters as an LLM-generated score, the standard errors reported below treat that regressor as error-free and so understate uncertainty.

\paragraph{Weighting and denominators.} Weighting follows each layer's native convention: the occupation-layer fit weights occupations by baseline posting volume, so the red line describes where demand mass moves, while the task-layer fit is unweighted across cells, so the blue line describes the grid itself. What the two gradients share is the sample, window, and metric; the weight is not among them. Two further asymmetries bound how the exact gap should be read. The denominators differ, postings at the occupation layer and mentions at the task layer, so a change in mentions per posting moves the blue line without any re-bundling; and exposure is itself a noisy score, whose attenuation shrinks with aggregation and therefore differs across layers. Swapped-weight estimates, an intensity decomposition of the mention-share change, and an attenuation-corrected slope are reported below.

\begin{figure}[t]
\centering
\includegraphics[width=0.84\textwidth]{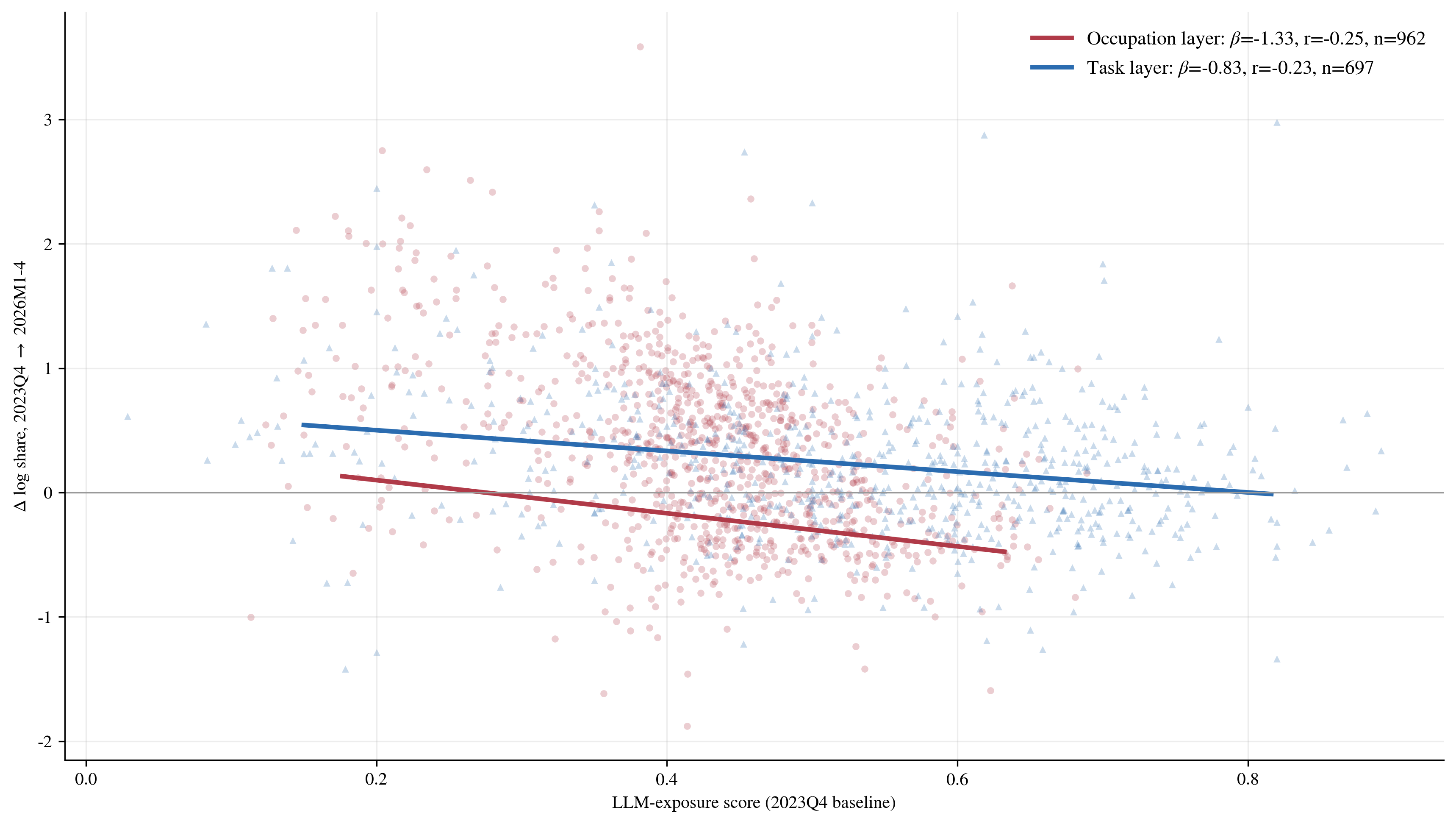}\\[6pt]
\includegraphics[width=0.66\textwidth]{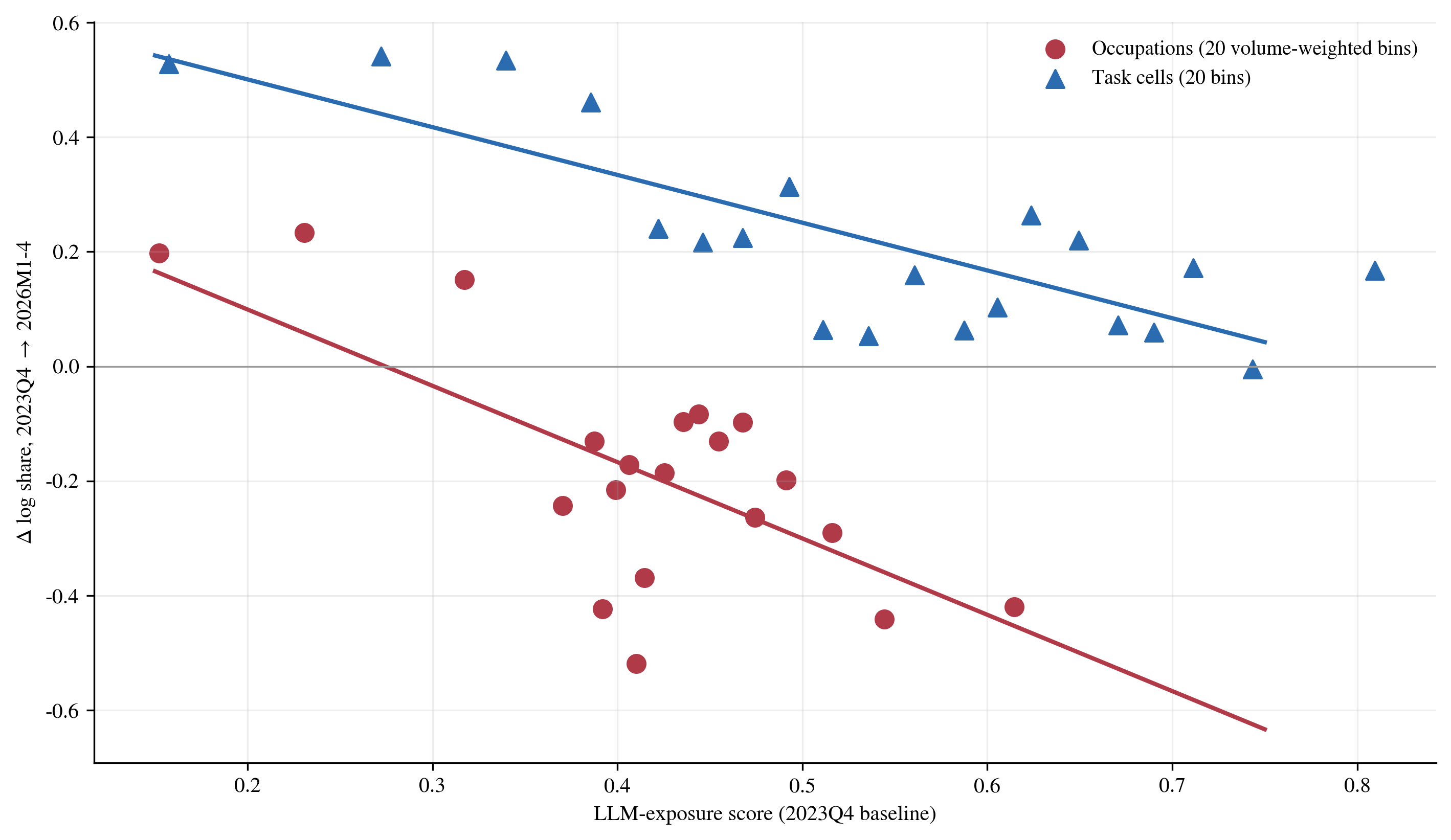}
\caption{Two related descriptive gradients in the same data. Horizontal axis: baseline (2023Q4) mean LLM-exposure score (0--1). Vertical axis: change in log share between 2023Q4 and January--April 2026, computed on the two largest platforms (A, B). \emph{Top panel}, red circles: each occupation ($n=962$ with at least 1{,}000 postings in both the 2023Q4 baseline and the January--April 2026 endline windows; exiting and entering occupations excluded), plotting the change in its log \emph{posting} share, with the volume-weighted fit ($\beta=-1.33$, $r=-0.25$). Blue triangles: each occupied cell of the $32\times26$ task grid ($n=697$), plotting the change in its log \emph{mention} share, with the unweighted fit ($\beta=-0.83$, $r=-0.23$). The two gradients share the exposure scale, window, and data; the outcome, its denominator, and the weighting differ, so the distance between the lines compares two descriptions rather than identifying a decomposition. High-exposure occupations lose posting share substantially faster than the high-exposure tasks inside them lose mention share; the shift-share and three-component decompositions reported in the text locate that gradient in the posting-share channel. \emph{Bottom panel}: the binned-scatter version on identical axes, with the 962 occupations grouped into twenty volume-weighted bins of baseline exposure and the 697 cells into twenty equal-count bins, against the full-sample fitted lines; each slope is a property of the whole distribution, not of outliers.}
\label{fig:twolayers}
\end{figure}

Figure~\ref{fig:twolayers} overlays the two layers. At the occupation layer the gradient is steep: $\beta=-1.33$ ($r=-0.25$, $n=962$). At the task layer it is markedly flatter: $\beta=-0.83$ ($r=-0.23$, $n=697$). Both slopes are in the same units, log-share change per unit of exposure score, so the two numbers sit on a common scale, though the comparison inherits the asymmetries noted above: for the same increment of exposure, occupations lose posting share substantially faster than tasks lose mention share. At either layer the correlation is modest ($r=-0.25$ and $-0.23$), so the slopes describe average tendencies across a widely dispersed distribution. High-exposure occupations are contracting faster on average in the demand pool; the high-exposure tasks inside them are declining far more slowly. How much of the adjustment is absorbed by the occupation mix rather than by the task content of surviving jobs is located by the decompositions below, not by this gap alone, since the two fits differ in outcome, denominator, and weighting. The occupation-layer gradient survives a same-season re-estimate under the identical rule (2024M1--4 $\to$ 2026M1--4, same platforms, same $\geq$1{,}000-posting filter in both windows, same weighting: $\beta=-1.46$, $r=-0.32$, $n=1{,}033$), so seasonal composition is not driving it. Neither slope is fragile: the occupation-layer estimate is $-1.33$ with cluster-robust standard error $0.46$ (clustered on the 41 top-level occupation groups of the working taxonomy), the task-layer estimate $-0.83$ with standard error $0.26$ clustered on the 32 business domains (heteroskedasticity-robust HC1 standard error $0.15$; $n=697$), both significant at the 1\% level under their layer's own weighting convention. The bottom panel of Figure~\ref{fig:twolayers} shows the binned-scatter version: twenty volume-weighted bins at the occupation layer, twenty equal-count bins at the task layer, against the full-sample fitted lines: each slope is a property of the whole distribution, not of outliers.

A shift-share decomposition locates the source of the gap. It is computed on the occupation $\times$ cell $\times$ period count matrix of Platform A, the largest platform in the two-platform pool, whose cell margins track the published task layer closely: the same 697 cells appear, and regressing the matrix's $\Delta$ log share on exposure returns $\beta=-0.835$ against the published two-platform $-0.834$ (recomputed on the identical two-platform matrix, the decomposition returns $\beta=-0.748$ with the same one-sided structure, composition $-0.695$ against re-bundling $-0.003$, so the allocation is not an artifact of the single-platform matrix). Each cell's share change splits exactly into a \emph{composition} term (occupations' posting weights shift, task mixes held fixed) and a \emph{re-bundling} term (occupations' own task mixes change, weights held fixed); the identity holds to machine precision, with the interaction absorbed symmetrically by the midpoint convention stated in Appendix~\ref{app:decomp}. Dividing each component by the cell's midpoint share puts it on the growth-rate scale of the figure ($\beta=-0.76$ for the recombined total, against $-0.83$ for the log difference), and each scaled component is then regressed on cell exposure with the same unweighted fit as the blue line. The result is one-sided: the composition term carries essentially the entire exposure gradient ($\beta=-0.74$, $r=-0.35$), while the re-bundling term is flat in exposure ($\beta=-0.01$, $r=-0.01$). On average, high-exposure tasks lose ground because the high-exposure \emph{occupations} that carry them shrink; within surviving occupations the estimated gradient is close to zero, so the data show no systematic stripping of exposed content from job descriptions. Task mentions decline more slowly than the occupations in which they were concentrated. Read this way, the flatter task-layer slope is largely the arithmetic consequence of composition: tasks are spread across occupations of varying exposure, which convolves the occupation-layer gradient into a flatter cell-layer one. The substantive finding is the near-zero re-bundling term, not the particular size of the slope gap.

\begin{figure}[t]
\centering
\includegraphics[width=0.58\textwidth]{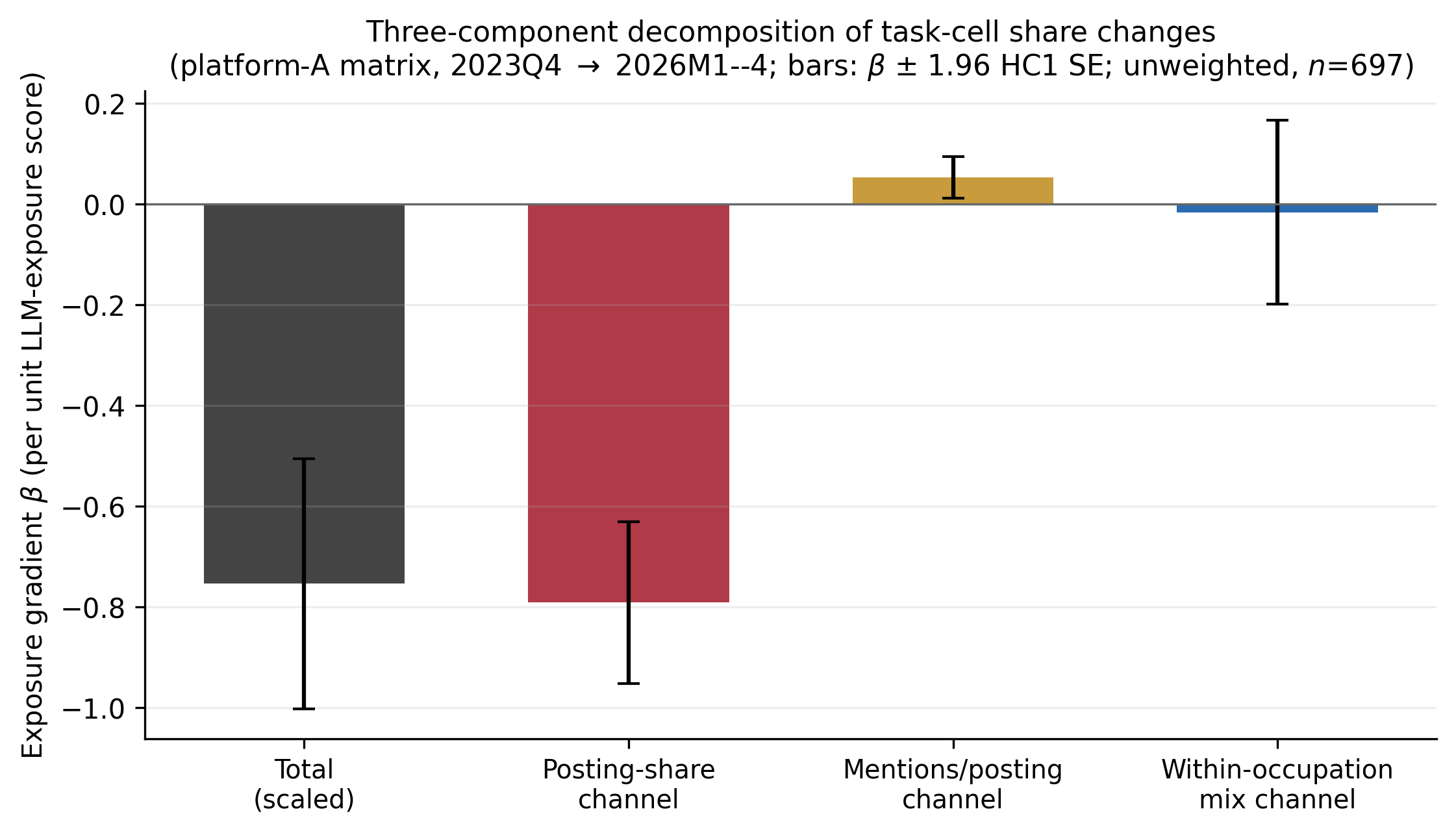}
\caption{Three-component decomposition of the task-layer share changes. Each cell's change in log mention share (2023Q4 $\to$ January--April 2026; computed on the occupation-by-cell matrix of Platform A, as in the shift-share decomposition; $n=697$) splits exactly into a posting-share channel (the cell's occupations gain or lose posting mass), a mentions-per-posting channel (postings carry more or fewer mentions), and a within-occupation mix channel (occupations' own task mixes change), allocated by the exact Shapley rule of Appendix~\ref{app:decomp}; bars report each channel's slope against cell LLM exposure, fitted as in Figure~\ref{fig:twolayers}. The gradient runs entirely through the posting-share channel ($\beta=-0.79$): high-exposure cells lose mention share because their occupations shrink, not because postings get shorter ($+0.05$) or because surviving jobs drop exposed content ($-0.02$).}
\label{fig:threecomp}
\end{figure}

\paragraph{Robustness.} Three exercises bound the two-layer comparison (scripts and full outputs released at \texttt{issue1\_results/}). First, the weighting asymmetry does not drive the gap: under all four weighting combinations the occupation-layer slope is far steeper than the task-layer slope, $-1.33$ against $-0.59$ when both layers are weighted by baseline volume, and $-2.94$ against $-0.83$ when both are unweighted, so the published mixed convention ($-1.33$ vs.\ $-0.83$) is the most conservative of the four. Second, the denominator asymmetry is silent. Decomposing each cell's mention-share change into a posting-share channel, a mentions-per-posting channel, and a within-occupation mix channel (three-factor Shapley decomposition, exact to machine precision; Figure~\ref{fig:threecomp}), the exposure gradient runs entirely through the posting-share channel ($\beta=-0.79$), the mentions-per-posting channel is if anything slightly positive ($\beta=+0.05$), which rejects the mechanical account in which high-exposure postings simply get shorter, and the within-occupation mix channel is flat ($-0.02$), consistent with the shift-share result above. Third, score noise does not drive the gap: the deployed grader's item-level signal share on LLM exposure is 0.91 (the share of its score variance that is signal, with true-score variance estimated from the fourteen-model exam's pairwise covariances; the exam's ICC(2,1) of 0.81 is the different quantity of a model drawn at random), the implied reliability of the cell-level exposure means is 0.971 (cell means average a median of eighteen items, shrinking item noise by the cells' weight concentration), and dividing by the cell-level factor moves the task-layer slope only from $-0.834$ to $-0.86$ (95\%~CI $-1.40$ to $-0.35$); the occupation layer, an average over far more mentions, attenuates even less. Inference is likewise robust: the occupation-layer estimate survives a wild cluster bootstrap over the 41 occupation groups ($p=0.008$), and the task-layer estimate survives clustering on the 32 business domains ($p=0.001$). The volume-weighted task-layer variant ($-0.59$) is only marginally significant under domain clustering ($p\approx0.08$--$0.10$), so we read the unweighted fit as the grid-level estimand.

An exploratory co-occurrence diagnostic, reported in Appendix~\ref{app:bundles}, asks where the surviving tasks appear: the fastest-emerging task pairings attach technical cores (troubleshooting, process optimization, equipment operations) to coordination, documentation, and compliance shells, at undiminished cognitive complexity. That material is exploratory and lives in Appendix~\ref{app:bundles}.

Together the two layers answer, within posted demand, the question of which layer moves: adjustment runs through which occupations are posted, not through what the surviving ones contain. Exposed occupations shrink, while the tasks they carried decline far more slowly, because those tasks also sit in occupations that are not shrinking. The tasks outlive the jobs that carried them. Both halves of that sentence are measured in the same data, at the same monthly frequency, with the same instrument, and both are descriptions: the pattern is as consistent with a demand cycle reshuffling occupations as with technology reshaping them. Persistence here is persistence at the grid's resolution: the gradients and decompositions live on the 697 domain-by-action cells, so what they establish is that broad task categories outlive the occupations that carried them, while churn among the 44{,}479 items inside a cell sits beneath this resolution.

\section{Versioning and limitations}
\label{sec:access}

\paragraph{Versioning.} The analysis scripts behind the figures and tables are committed with the paper source (\texttt{code/}), alongside the per-occupation validity inputs (\texttt{figures/occ\_llm\_exposure.csv}). Because a full re-scan of the data under a fixed catalog completes in hours, the catalog is allowed to evolve, and versioning is what makes an evolving instrument citable. Each release is an immutable snapshot identified by a catalog version and a score version; when the taxonomy is re-derived or the grader rubric revised, the full history is re-scanned under the new version, the new snapshot is archived alongside the old, and published results remain reproducible against the vintage they cite. The re-runnable design thus yields a vintage policy rather than a moving target: ``measured under catalog v$X$, scores v$Y$, data vintage $Z$'' is a complete specification of any number in this paper. For this paper itself, every statistic and figure is measured under catalog v2026-04-04 (the fixed master of 65{,}200 items, 20{,}721 requirement and 44{,}479 task buckets, compiled to 353{,}359 patterns) and scores v2026-04 (the deployed grader's single batch over that master, 65{,}200 calls). Figures and regressions are computed from the 2026-04-26 rollup vintage, whose analysis windows close in April 2026, the last complete month at computation time; the posting totals of Figure~\ref{fig:corpus} are measured under data vintage v2026-06, the 2026-04-26 rollup base plus incremental batches 202604 and 202606, each recorded with source fingerprints in a machine-readable apply ledger. A third batch, 202607, was applied before the figures were last re-rendered, so re-rendered figures carry its retroactive revisions; the visible movements are at the scale of one percentage point, the accountant task core moving from 69 to 68 per cent being the example in this paper. The resource footprint behind the efficiency claim is one workstation, an Intel Core i9-13950HX with 224~GB of RAM, on which a full Hyperscan re-scan of the 116~million unique descriptions completes in hours; the entire scoring bill for the catalog was \yen5{,}473 and the boundary-adjudication bill about \yen3{,}300, both one-off artifact costs independent of the size of the data.

\paragraph{Limitations.} Five are structural. Postings over-represent urban, formal, white-collar and service hiring; Table~\ref{tab:representativeness} bounds that tilt but does not remove it. A posting is an advertisement, neither a vacancy count nor a hire, so the system measures expressed demand composition, not employment. The regex backfill prefers precision over recall, so mention counts are conservative, and rare phrasings are under-captured. The score dimensions are judged at current-model capability: LLM exposure, in particular, is a statement about what today's models absorb, frozen at scoring time. That is why scores are versioned and re-scorable rather than permanent, and why downstream causal work should anchor on realized outcomes (Section~\ref{sec:twolayers}) rather than on scores alone, bringing its own identification: the gradients there are descriptive. And the validation program is machine-benchmarked end to end: extraction and backfill are checked against machine-generated references, and scores are compared against a model-consensus exam, so the reported agreement establishes reproducibility relative to the recorded contracts rather than human-judged accuracy.

\section{Conclusion}
\label{sec:conclusion}

Occupational titles are informative but incomplete descriptions of posted work. To measure what titles leave hidden, this paper introduces a dynamic, posting-based occupational information system for China, constructed from 752.6~million job advertisements. It organizes employer-written text into 20{,}721 candidate-requirement items and 44{,}479 advertised-task items. Matching these items back to every posting yields 12.1~billion mentions that can be tracked monthly by occupation, city, wage, education, and employer type. Across the two catalogs, thirteen continuous dimensions are defined---eleven for tasks and nine for requirements. The extraction and backfill are benchmarked bucket by bucket against machine-annotated references, and the resulting occupation-level LLM-exposure measure correlates at $r=0.73$--$0.79$ with leading published exposure indices. Once the vocabulary is fixed, the full history re-scans in hours. Methodologically, the system shows that a small team can build O*NET-scale infrastructure from raw market text by spending model judgment only on reusable artifacts, taxonomies, boundaries, crosswalks, and patterns, and replaying them over arbitrarily large corpora with embeddings and compiled automata.

The two examples show what the change of instrument buys. The first reveals what a single title hides on both sides at once: task and requirement profiles diverge sharply within some occupations and remain stable within others, as the contrast between accountants and Java developers shows. Among accountant postings, higher pay is associated with a shift from core bookkeeping toward compliance and data analysis, while the credential mix moves from the junior to the intermediate accounting qualification. The second example shows why the level of measurement matters for reading change over time. From late 2023 onward, posted work shifted toward more hands-on, on-site, and higher-stakes content, but fixed-composition counterfactuals attribute most of that movement to re-weighting across occupations. More LLM-exposed occupations lost posting share faster than the broad task categories associated with them, and the decompositions locate the exposure gradient in the posting-share channel, with the within-occupation task-mix channel approximately flat. Adjustment-cost logic motivated the possibility that employers might recompose work before headcounts move; what this window shows is narrower but more precise---the observable adjustment in posted demand ran primarily through the occupation mix.

Taken together, the central lesson is that the interpretation of labor-market change depends on the level at which it is measured. Occupational titles are useful classifications, but they are not sufficient statistics for posted work: the task bundles they contain and the entry requirements associated with them can vary across postings and need not move together. These findings concern online posted demand rather than hires or employment, and they do not identify whether the observed changes were caused by AI or whether they lead official employment counts. The record's monthly structure nevertheless enables sustained tracking of advertised demand, task re-bundling across the occupational space, new-work detection \citep{autor2024newfrontiers}, and the mapping of education requirements to task content. Understanding technological change therefore requires tracking not only which occupational titles expand or contract, but also which tasks and entry requirements persist beneath them.

\clearpage

\appendix
\section{Exploratory diagnostic: co-occurrence bundles}
\label{app:bundles}

\begin{figure}[t]
\centering
\includegraphics[width=0.85\textwidth]{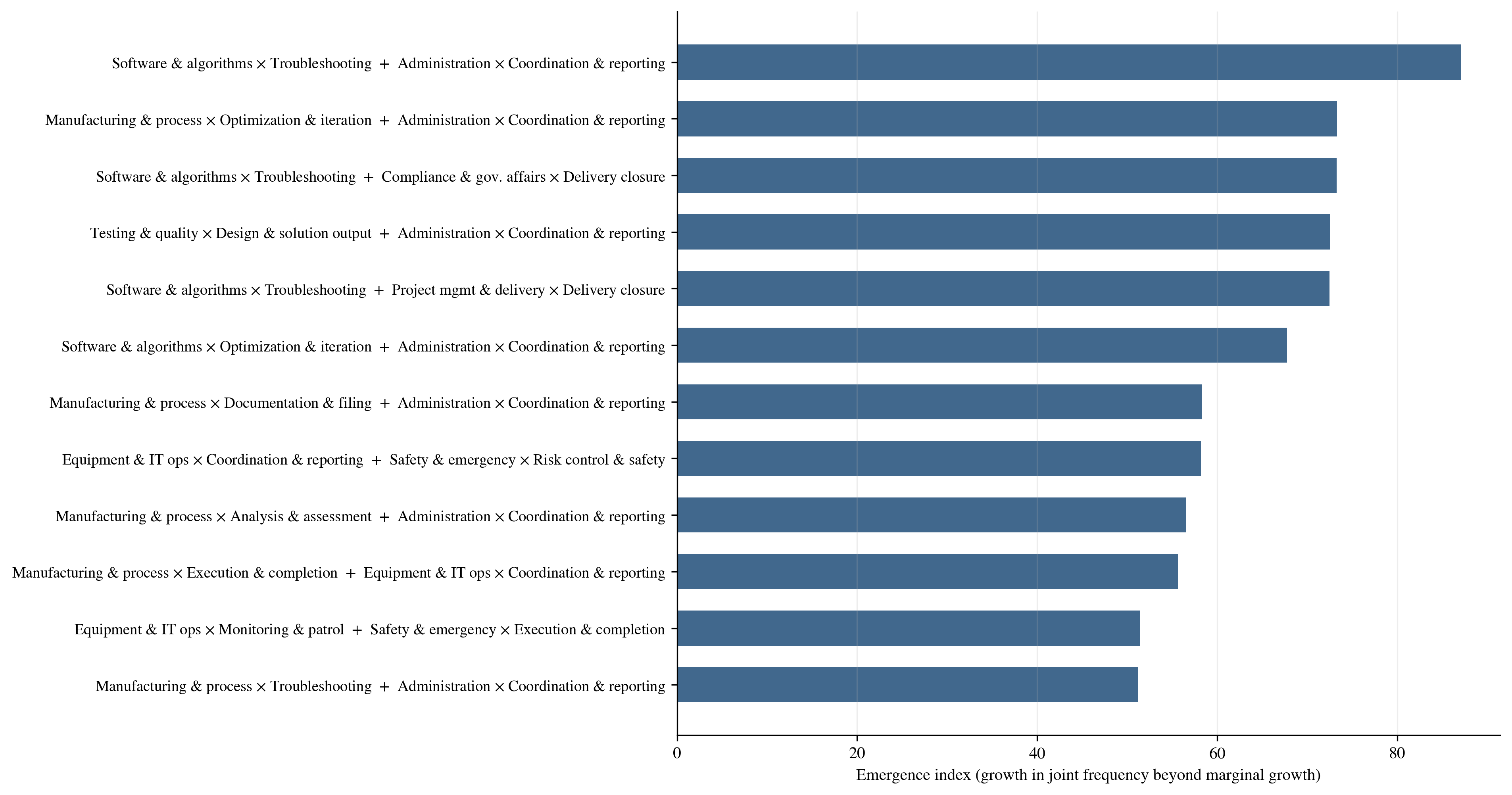}
\caption{The twelve fastest-emerging task bundles, 2023Q4 $\to$ January--April 2026 (two largest platforms). A bundle is a pair of task-grid cells (each labeled \emph{business domain} $\times$ \emph{action}) whose mentions co-occur in the same posting. The index combines volume, growth, excess co-occurrence over the two cells' margins, and cross-occupation spread, penalized by baseline size (the components are stated in the text above), so small-base, fast-spreading pairings rank highest. The grammar of the fastest-emerging combinations is uniform: a technical core (software troubleshooting, manufacturing-process optimization, equipment operations) joined to a coordination, documentation, or compliance shell (administrative coordination \& reporting, delivery closure, risk control \& safety). This is a descriptive pattern.}
\label{fig:newbundles}
\end{figure}

This appendix reports the co-occurrence diagnostic referenced in Section~\ref{sec:twolayers}. We call a pair of task-grid cells a \emph{bundle} when the two co-occur in the same posting, and score each bundle by an emergence index, an equal-weight sum of five standardized components: current joint volume, growth in joint frequency, growth in pointwise mutual information over the two cells' margins, growth in the number of occupations carrying the pair, and, entering negatively, baseline volume, so that small-base, fast-growing, increasingly cross-occupation pairings rank highest. Co-occurrence is counted per posting, capped at thirty cells per posting, and a pair without baseline support enters the mutual-information term at zero. The index rewards excess co-occurrence against these benchmarks; it does not by itself certify a combination as new work. Figure~\ref{fig:newbundles} lists the twelve fastest-emerging bundles. Their grammar is uniform: a technical core joined to a coordination, documentation, or compliance shell. Software troubleshooting arrives bundled with administrative coordination and reporting, a combination spreading through quality-assurance, developer, project-management, and customer-service postings; manufacturing-process optimization, analysis, and documentation each arrive bundled with the same coordination shell; equipment operations arrives bundled with safety and risk response, spreading through security, production, and facility roles. Read descriptively, these pairings are consistent with widening job scopes, with technical content increasingly arriving together with coordination, documentation, and compliance content in the same posting. We stress the two caveats that keep this material out of the main text: coordination-and-reporting language is among the most formulaic in job advertisements, and co-occurrence alone cannot distinguish genuine task reorganization from boilerplate inflation. The diagnostics here are descriptive screening devices, not evidence of job redesign.

\begin{figure}[t]
\centering
\includegraphics[width=0.78\textwidth]{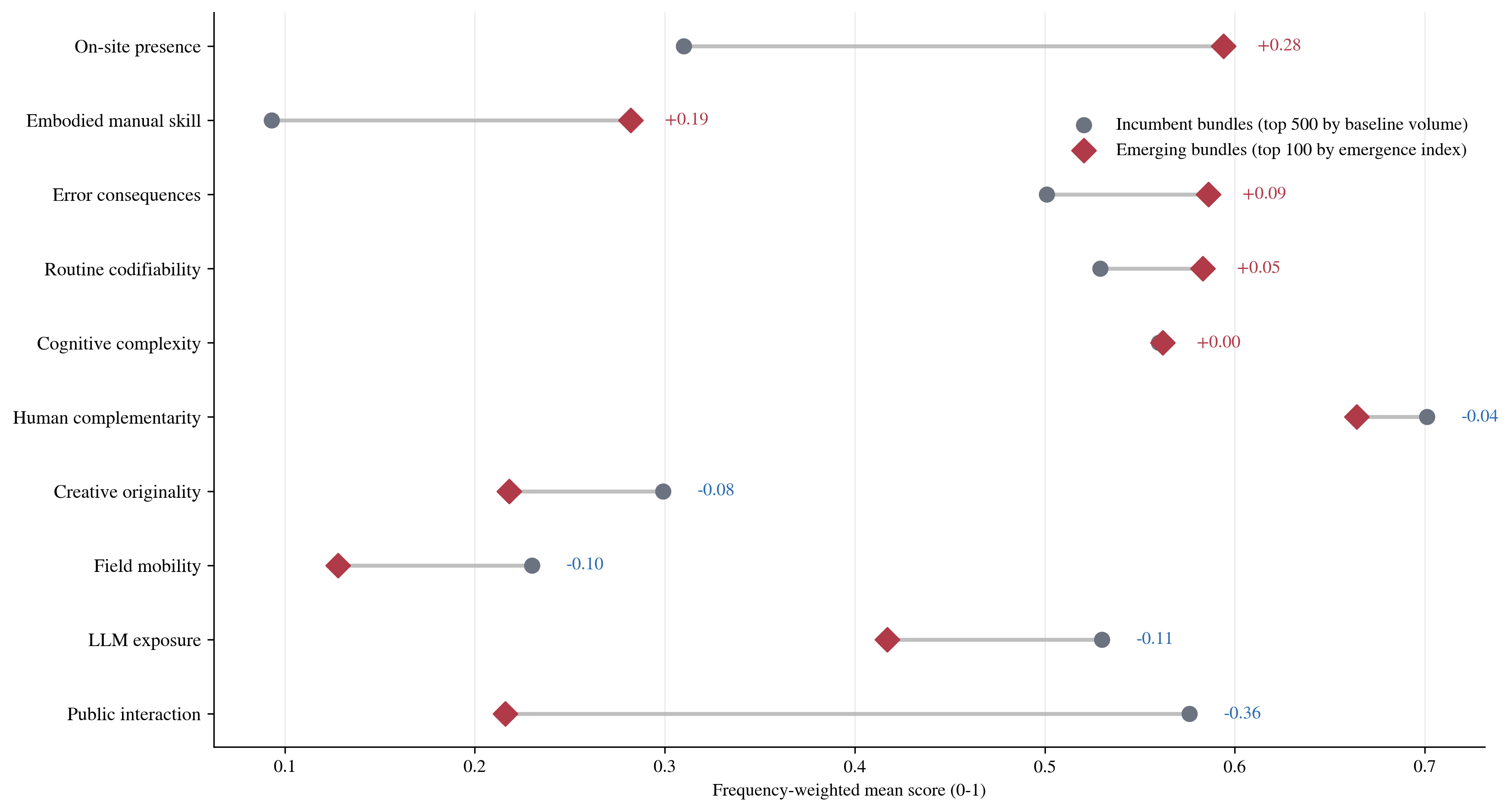}
\caption{What the emerging bundles are made of. Each row compares two sets of task bundles on one 0--1 dimension: the 100 fastest-emerging bundles (red diamonds; ranked by the emergence index of Figure~\ref{fig:newbundles}) versus the 500 highest-volume incumbent bundles at the 2023Q4 baseline (gray circles). Points are frequency-weighted mean scores; labels give the difference (emerging minus incumbent). The emerging combinations are sharply more on-site ($+0.28$) and more embodied ($+0.19$), carry higher error consequences ($+0.09$), and are far less public-facing ($-0.36$) and less LLM-exposed ($-0.11$), at identical cognitive complexity ($+0.00$) and nearly identical human complementarity ($-0.04$).}
\label{fig:bundledims}
\end{figure}

Figure~\ref{fig:bundledims} compares the emerging bundles (top 100 by emergence index) with the incumbent high-volume bundles (top 500 by baseline volume) across ten dimensions, frequency-weighted. The emerging combinations are sharply more on-site (+0.28) and more embodied (+0.19), carry higher error consequences (+0.09), and are far \emph{less} public-facing ($-0.36$) and less LLM-exposed ($-0.11$), yet their cognitive complexity is identical (+0.00) and their human complementarity nearly so ($-0.04$). The high-wage production-worker ads of Example~1, assembly titles describing documentation, data recording, and cross-department coordination, are consistent with such bundles appearing inside a single occupation.

\clearpage

\section{Models behind the judgment stages}
\label{app:models}

Table~\ref{tab:models} records the model behind every LLM-marked node of the stage diagrams, as logged in the run configurations, together with the embedding stack. Several models appear because the choices were made stage by stage: where a judgment shapes results, the model was selected by benchmark, the deployed grader by the fourteen-model exam of Section~\ref{sec:scores} and the regular expressions by keeping the best of five generators per bucket, since no single generator wins universally (Section~\ref{sec:backfill}); elsewhere the model of record is documented so that every judgment can be traced to the engine that made it.

\begin{table}[t]
\centering
\caption{Models behind the construction's LLM-judgment stages, as recorded in run configurations; the industry-vocabulary crosswalk of \S\ref{sec:ind} predates this logging. Data-scale processing uses no LLM: cached lookup tables, embeddings, and compiled regular expressions do that work.}
\label{tab:models}
\footnotesize
\begin{tabularx}{\textwidth}{lX}
\toprule
\textbf{Stage} & \textbf{Model(s)} \\
\midrule
Occupation-category crosswalk (\S\ref{sec:occ}) & \texttt{claude-opus-4-6} (batch) \\
Closed-book coding benchmark (rejected design, \S\ref{sec:occ}) & \texttt{gpt-5.5}, \texttt{claude-opus-4-7}, \texttt{deepseek-v4-pro}, \texttt{qwen3.6-plus}, \texttt{doubao-seed-2-0-pro}, \texttt{glm-5.1} \\
Embedding (title backfill, \S\ref{sec:occ}; candidate recall, \S\ref{sec:reqres}) & \texttt{Qwen3-Embedding-0.6B}, self-hosted; vectors reduced to 64 dimensions by principal component analysis (PCA); inverted-file (IVF) index \\
Major surface-form dictionaries (\S\ref{sec:major}) & \texttt{claude-3-5-sonnet} (2024-12 build, carried forward) \\
Atomic extraction (\S\ref{sec:reqres}) & pilot \texttt{deepseek-reasoner} (2026-01); production \texttt{doubao-seed-2-0-lite-260215} over the 2.31M-job-description (JD) sample \\
Boundary adjudication (\S\ref{sec:reqres}) & \texttt{claude-opus-4-6} (batch, $\sim$520{,}000 pairs) \\
Grader exam (\S\ref{sec:scores}) & fourteen candidates: \texttt{claude-opus-4-6}, \texttt{claude-opus-4-7}, \texttt{gpt-5.5}, \texttt{gpt-5.4}, \texttt{gemini-3.1-pro-preview}, \texttt{deepseek-v4-pro}, \texttt{deepseek-v4-flash}, \texttt{doubao-seed-2-0-pro-260215}, \texttt{doubao-seed-2-0-lite-260215}, \texttt{kimi-k2.5}, \texttt{kimi-k2.6}, \texttt{glm-5.1}, \texttt{qwen3.6-plus}, \texttt{hy3-preview} \\
Deployed grader (\S\ref{sec:scores}) & \texttt{gpt-5.5} (batch, high reasoning effort; 65{,}200 items) \\
Regex generation, best-of-5 (\S\ref{sec:backfill}) & \texttt{gpt-5.4}, \texttt{claude-opus-4-6}, \texttt{doubao-seed-2-0-pro-260215}, \texttt{doubao-seed-2-0-lite-260215}, \texttt{gemini-3-flash-preview} \\
\bottomrule
\end{tabularx}
\end{table}

\section{Decomposition definitions}
\label{app:decomp}

Both decompositions of Section~\ref{sec:twolayers} operate on the occupation-by-cell count matrix of Platform~A, $n_{oc}^{t}$, where $o$ indexes occupations, $c$ the 697 occupied cells of the $32\times26$ task grid, and $t\in\{0,1\}$ the windows 2023Q4 and January--April 2026.

\paragraph{Two-term split.} Let $T^{t}=\sum_{o,c}n_{oc}^{t}$, let $w_{o}^{t}=\sum_{c}n_{oc}^{t}/T^{t}$ be occupation $o$'s share of all task mentions, and let $m_{oc}^{t}=n_{oc}^{t}/\sum_{c}n_{oc}^{t}$ be its within-occupation cell mix. The cell's mention share is $s_{c}^{t}=\sum_{o}w_{o}^{t}m_{oc}^{t}$. Writing $\bar{x}=(x^{0}+x^{1})/2$ and $\Delta x=x^{1}-x^{0}$,
\[
\Delta s_{c}\;=\;\underbrace{\sum_{o}\Delta w_{o}\,\bar{m}_{oc}}_{\text{composition}}\;+\;\underbrace{\sum_{o}\bar{w}_{o}\,\Delta m_{oc}}_{\text{re-bundling}},
\]
an identity with no residual: evaluating each channel at the midpoint of the other absorbs the interaction symmetrically, so no convention beyond the midpoint is needed. Each term is divided by the midpoint share $\bar{s}_{c}$ to put it on a growth-rate scale (for small changes $\Delta s_{c}/\bar{s}_{c}\approx\Delta\log s_{c}$, which is why the scaled total fits exposure at $-0.76$ against $-0.83$ for the log difference), and each scaled term is regressed on cell exposure, unweighted, with HC1 standard errors.

\paragraph{Three-component split.} The three-component decomposition of Figure~\ref{fig:threecomp} works on posting-based factors: occupation posting shares $p_{o}^{t}$, mentions per posting $\nu_{o}^{t}$, and the mix $m_{oc}^{t}$, with $s_{c}(\theta)\propto\sum_{o}p_{o}^{\theta_{1}}\nu_{o}^{\theta_{2}}m_{oc}^{\theta_{3}}$ for a switch vector $\theta\in\{0,1\}^{3}$. Each factor group's contribution is its exact Shapley value: the average, over all $3!$ orderings of switching the three groups from $t=0$ to $t=1$, of the change in $s_{c}$ when that group switches. The three contributions sum to $s_{c}^{1}-s_{c}^{0}$ to machine precision, are scaled by the midpoint share as above, and are fitted to exposure by the same rule. The sample keeps occupations with positive postings and mentions in both windows.

\section*{Disclosures}
The raw posting archive is used under data-use agreements with the platforms, which permit release of the catalog, scores, crosswalks, and aggregated statistics but not raw ad text, posting-level microdata, or employer-identifiable information; platform identities are anonymized throughout in accordance with those agreements. Qin Chen is affiliated with Shanghai DeepPulse Technology. Ying Fang acknowledges the support of the National Natural Science Foundation of China (Project 72595870; 72595875). The study analyzes commercially posted job advertisements in aggregate; no individual-level personal data are redistributed.

\bibliographystyle{plainnat}
\bibliography{references}

\end{document}